\documentclass[fleqn,usenatbib]{mnras}

\usepackage{newtxtext,newtxmath}

\usepackage[T1]{fontenc}
\usepackage{bm}
\usepackage{booktabs}
\usepackage{array}
\usepackage{enumerate}
\usepackage{ulem}
\usepackage{threeparttable}
\usepackage{multirow}
\usepackage{soul}
\setstcolor{red}
\usepackage{color}
\newcommand{\add}[1]{#1}
\DeclareRobustCommand{\VAN}[3]{#2}
\let\VANthebibliography\thebibliography
\def\thebibliography{\DeclareRobustCommand{\VAN}[3]{##3}\VANthebibliography}

\usepackage{graphicx}	
\usepackage{amsmath}	

\title[Electromagnetic Follow-ups of Binary Neutron Star Mergers]{Electromagnetic \add{Follow-up Observations} of Binary Neutron Star Mergers with Early Warnings from Decihertz Gravitational-wave Observatories}

\author[Y. Kang, et al.]{
Yacheng Kang,$^{1,2}$
Chang Liu,$^{1,2}$
and Lijing Shao$^{2,3}$\thanks{E-mail: lshao@pku.edu.cn}
\\
$^{1}$Department of Astronomy, School of Physics, Peking University, Beijing 100871, China\\
$^{2}$Kavli Institute for Astronomy and Astrophysics, Peking University, Beijing 100871, China\\
$^{3}$National Astronomical Observatories, Chinese Academy of Sciences, Beijing 100012, China}

\date{Accepted XXX. Received YYY; in original form ZZZ}

\pubyear{2022}

\begin{document}
\label{firstpage}
\pagerange{\pageref{firstpage}--\pageref{lastpage}}
\maketitle

\begin{abstract}
We investigate the prospects of electromagnetic \add{follow-up observations} for binary neutron
star (BNS) mergers, with the help of early warnings from  decihertz
gravitational-wave (GW) observatories, B-DECIGO and DO-Optimal. \add{Extending the
previous work,} we not only give quick assessments of
joint short $\gamma$-ray burst (sGRB) detection rates for different $\gamma$-ray
satellites and BNS population models, but also elaborate on the analyses and
results on multi-band kilonova detections for survey telescopes with different
limiting magnitudes. During an assumed 4-year mission time for decihertz GW
observatories, we find that for the goals of electromagnetic follow-ups,
DO-Optimal performs better than B-DECIGO as a whole on the detection rate, and
has a larger detectable distance for joint sGRB/kilonova searches. Taking the
log-normal population model for BNS mergers and a $\text{one-day}$ early-warning
time as an example, we discuss the accuracy in localization and timing, as well
as the redshift distributions for various synergy observations with
electromagnetic facilities and decihertz GW detectors.  Based on our analyses,
we propose a feasible ``wait-for'' pattern as a novel detecting mode for future
multi-messenger astrophysics.
\end{abstract}




\section{Introduction}
\label{ sec:intro }

The first binary neutron star (BNS) merger, GW170817, was initially alerted by
the advanced LIGO and Virgo detectors \citep{LIGOScientific:2017vwq}, and the
gravitational wave (GW) signal was followed up by  intensive observations across
the whole electromagnetic (EM) spectrum \citep{LIGOScientific:2017ync}. Because
of this, GW170817 marks a spectacular success of multi-messenger astrophysics,
and has  been viewed as the watershed moment in astronomy and physics
\citep{Kalogera:2021bya}.  From the combination of GW and EM data, its
luminosity distance is determined to be $D_{\mathrm{L}} \approx
40\,\mathrm{Mpc}$ \citep{Hjorth:2017yza}.  Thanks to the short distance,
GW170817 has been subsequently confirmed in association with a short
$\gamma$-ray burst (sGRB)  $\sim 1.7\,\mathrm{s}$ after the merger
\citep[GRB\,170817A;][]{LIGOScientific:2017zic, Goldstein:2017mmi,
Savchenko:2017ffs, 2018NatCo...9..447Z},  a kilonova discovered in the galaxy
NGC\,4993 $\sim 11\,\mathrm{hr}$ later \citep[AT 2017gfo;][]{Coulter:2017wya,
Evans:2017mmy, Pian:2017gtc, Kilpatrick:2017mhz}, and an off-axis
multi-wavelength afterglow detected with its luminosity peaking at $\sim
200\,\mathrm{d}$ \citep{Margutti:2017cjl, Troja:2017nqp, Lazzati:2017zsj,
Lyman:2018qjg, Ghirlanda:2018uyx}. Such a multi-messenger observation has not
only provided us the smoking-gun evidence for the BNS merger origin of sGRBs and
kilonovae \citep{LIGOScientific:2017zic, Li:1998bw}, but also an unprecedented
opportunity to explore dense matter properties and gravity theories in extreme
environments \citep{LIGOScientific:2017pwl, LIGOScientific:2018cki,
LIGOScientific:2018dkp}. These aspects are vital to many open questions in the
fundamental physics \citep{Sathyaprakash:2019yqt, Arun:2022vqj}.

However, in addition to GW170817, there is only one other potentially heavy BNS
merger  event to date in the LIGO/Virgo/KAGRA data
\citep[GW190425;][]{LIGOScientific:2020aai}. The fast-evolving nature of sGRBs
and kilonovae might appear discouraging for multi-messenger observations after
the BNS mergers \citep{Zhu:2021zmy}, let alone the common faintness for current
EM follow-up facilities due to the large distances between most BNS mergers and
the Earth \citep{Zhu:2021ram}. Owing to many difficulties for all-sky
multi-messenger detections of BNS mergers using the traditional time-domain
survey projects, some studies have been devoted to the best searching strategy
with ground-based GW triggers for EM follow-up observations
\citep{Cowperthwaite:2015kya, Gehrels:2015uga, Rosswog:2016dhy,
Cowperthwaite:2018gmx, Liu:2020bgc}, especially focusing on kilonova detections by
serendipitous observations \citep{Metzger:2011bv, Coughlin:2017ydf,
Coughlin:2020fwx, Zhu:2020ffa, Zhu:2021ram}. What were discussed in these
studies have included different proposals for present and future GW/EM synergy
projects,  characterized by the field of view (FoV), search cadence, filters,
exposure time, and so on.  The complexity of these schemes can be overwhelming.

Differently from these former investigations, in this work we discuss the
prospects for the multi-messenger observations with GW early warnings using
space-borne decihertz GW detectors. As \citet{Liu:2022mcd} recently
demonstrated, with decihertz GW detectors alone, the localization and timing
accuracy of BNS mergers may achieve a level of $
\mathcal{O}(0.01)\,\mathrm{deg}^2$ and $ \mathcal{O}(0.1)\,\mathrm{s}$,
respectively. We propose that such results will allow us to prepare a realistic
``wait-for'' pattern for various $\gamma$-ray satellites and survey telescopes
with current or future designs, even in the case that we  request a
$\text{one-day}$ early-warning time before the BNS merger. By contrast, current
and future ground-based GW detectors can not provide localization estimates as
accurately as the space-borne decihertz GW detectors \citep{Magee:2022kkc,
Borhanian:2022czq}. Due to the higher sensitive frequency range of the
ground-based detectors, they can not offer alerts as early as the decihertz GW
detectors, either. Without a sufficient early-warning time, it is also hard for
EM telescopes to capture the early evolution of each detectable BNS merger, even
for the low-latency GW-trigger events.  EM follow-up facilities usually need
response time to make decisions and tune satellites and telescopes.
Nevertheless, our results on the EM \add{follow-up observations} of BNS mergers with space-borne
decihertz GW detectors will bring excellent opportunities for future
multi-messenger astronomy.

We study a wait-for scheme with  four BNS population models and two
representative decihertz GW detectors, B-DECIGO
\citep[B-DEC;][]{Isoyama:2018rjb, Kawamura:2020pcg} and DO-Optimal
\citep[DO-OPT;][]{Sedda:2019uro}. Taking the log-normal population model for BNS
mergers and a $\text{one-day}$ early-warning time as an example, our results
show that both B-DEC and DO-OPT can reach a localization accuracy of $\Delta
\Omega \lesssim 1\,\mathrm{deg}^2$ and timing accuracy of $\Delta t_{\rm c}
\lesssim 0.5\,\mathrm{s}$. This will be of great use to synergy observations
with EM facilities. We also find that DO-OPT is expected to have a better
performance than B-DEC as a whole on the joint sGRB/kilonova detection rate,
especially for the high-redshift events. With the redshift distributions for
sGRB/kilonova searches presented in this work, we provide meaningful references
and helpful inputs for upcoming EM follow-up projects. We admit that a specific
ground-based optical survey mission should consider the discount due to its
maximum detectable sky coverage and the Earth's rotation. But this can be
partially solved if there are at least two survey telescopes on the different
sides of the Earth. More importantly, due to the accurate localization and
timing ability of decihertz GW detectors with a sufficient early-warning time in
our scenario, there is no need to consider a complex searching strategy for the
EM follow-up observations anymore. Such a wait-for pattern could also provide a rare opportunity for analyzing the early evolution
characteristics of each detectable system, which we leave to
future studies.

The organization of this paper is as follows. Following the method and procedure
presented in \citet{Liu:2022mcd}, in Sec.~\ref{ sec:EARLY-WARNING POPULATION }
we first overview the BNS merger populations and the GW detecting strategy with
two space-borne decihertz GW observatories. For EM follow-up observations, we
present our methodologies on the sGRB and kilonova detections in Sec.~\ref{
sec:EM Counterpart Detection }. Using the above ingredients, we report detailed
analyses and results on various aspects of multi-messenger early-warning
detections in Sec.~\ref{ sec:Results }. Finally, Sec.~\ref{ sec:Conclusion }
concludes the paper. We use geometric units where $G=c=1$.

\section{ GW Early Warnings of BNS Mergers} 
\label{ sec:EARLY-WARNING POPULATION }

As mentioned in the Introduction, one major difference between our work and
other researches on multi-messenger observations of BNS mergers is that we aim
to explore how to achieve the wait-for pattern for various EM facilities with
the early warnings from decihertz GW detectors. Therefore, in this section, we
briefly overview the early-warning populations for later analyses. 

\begin{figure}
    \centering
    \includegraphics[scale=0.45]{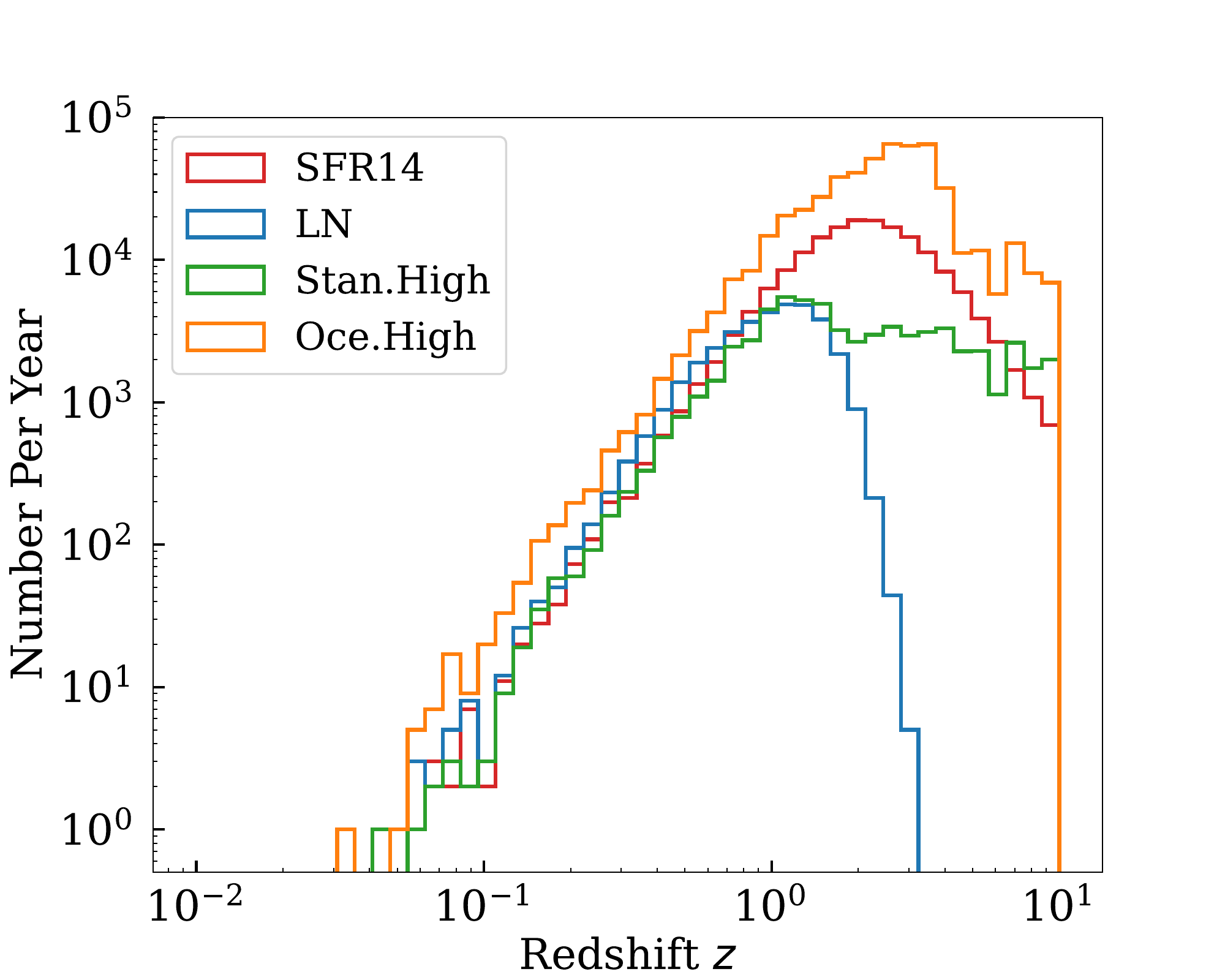}
    \caption{Total number of BNS mergers per year for different population
    models.} \label{ fig:population }
\end{figure}

Before quantitatively analyzing the early warnings from GW observatories and the
EM counterparts from BNS mergers, one needs to obtain their redshift
distributions in our Universe. Given that there are various types of population
models in  literature and their predicted results vary widely among different
models, especially for the high-redshift regime, we follow \citet{Liu:2022mcd}
and consider four kinds of population models. The models---abbreviated as
``SFR14'', ``LN'', ``Stan.High'', and ``Oce.High''---are briefly summarized as
follows:
\begin{enumerate}[(1)]
   \item \textbf{SFR14}---BNS merger rate evolves with redshift following the
   fitting formula of star formation rate (SFR) in \citet{Madau:2014bja};
   \item \textbf{LN}---We consider the log-normal delay model based on a delay
   time superposed on the SFR \citep{Wanderman:2014eza, Sun:2015bda,
   Zhu:2020ffa};
   \item \textbf{Stan.High}---We adopt the standard model with high-end
   metallicity evolution scenarios from \citet{Dominik:2013tma};
   \item \textbf{Oce.High}---We adopt the optimistic common envelope model with
   high-end metallicity evolution scenarios from \citet{Dominik:2013tma}.
\end{enumerate}
Based on the  GWTC-3 PDB (ind) model \citep{LIGOScientific:2021psn}, we adopt
the local merger rate $44\,\mathrm{Gpc}^{-3}\,\mathrm{yr}^{-1}$ for BNSs. The
redshift cutoff is at $z\simeq10$ and more details about BNS  populations can be
found in \citet{Liu:2022mcd}. Note that a standard $\Lambda \mathrm{CDM}$
cosmology with $H_{0}=67.8\,\mathrm{~km}\,\mathrm{~s}^{-1}\,\mathrm{Mpc}^{-1}$,
$\Omega_{\Lambda}=0.692$, and $\Omega_{\mathrm{m}}=0.308$ is assumed in this
work  \citep{Planck:2015fie}.  Finally, we plot the distributions of BNS mergers
in Fig.~\ref{ fig:population } for the above population models.\footnote{\add{In
a recent paper, \citet{OConnor:2022lpb} have suggested that a power-law delay
time distribution with a decay index between $\sim -1$ and $-1.5$ seems to be
consistent with the deeper observational results. If this is confirmed, we
caution that the LN model might underestimate the rate of high-$z$ events.}}

For each BNS, their NS masses, $M_{1}$ and $M_{2}$, are randomly generated based
on the observational distribution of Galactic BNS systems, using a normal
distribution $M_{1,2} / M_{\odot} \sim \mathcal{N}\left(1.32,0.11^{2}\right)$
for each NS \citep{Lattimer:2012nd, Kiziltan:2013oja}.  This distribution has
been adopted in other studies as well \citep{Song:2019ddw, Yu:2021nvx,
Zhu:2021ram}. With known $M_{1,2}$ and redshift $z$, we will discuss the
prospects for multi-messenger early-warning detections of BNS mergers in the
following sections.

For multi-messenger detections of BNS mergers, the space-borne decihertz GW
detectors could offer alerts much earlier than ground GW detectors and EM
facilities. We report detailed analyses with early warnings only from B-DEC and
DO-OPT for in-depth comparisons. Conservatively, we set a 4-yr mission time for
the two decihertz GW detectors in our work. More detailed descriptions of these
detectors are presented in, e.g., \citet{Liu:2021dcr}.

Recently, \citet{Liu:2022mcd} have classified all BNS mergers in decihertz GW
detectors into 3 categories based on the observational properties,  in
particular the accuracy of angular resolution $\Delta \Omega$ and time of merger
$\Delta t_{\mathrm{c}}$. Three categories are:
\begin{enumerate}[(a)]
  \item BNSs that merge within 1 year since the start of observation
  ($t_{\mathrm{c}_{0}} \leq 1\,\mathrm{yr}$);
  \item BNSs that merge in 1 to 4 years ($1\,\mathrm{yr}<t_{\mathrm{c}_{0}}
  \leq 4\,\mathrm{yr}$);
  \item BNSs that only inspiral within the whole 4-yr observational span
  ($t_{\mathrm{c}_{0}}>4\,\mathrm{yr}$).
\end{enumerate}
Here $t_{\mathrm{c}_{0}}$ is defined as the merger time since the mission
begins. For sources in category (a), due to their short stay in the decihertz
detector, not enough information could be accumulated to obtain precise
parameter estimations. As an example, the angular resolution for B-DEC could
reach ${\cal O}(10^4)\,\mathrm{deg}^2$, which may not be covered by the FoV of
EM telescopes, let alone for the wait-for pattern discussed in this work. For
mergers in category (c), although space-borne decihertz detectors could provide
warnings much earlier than the ground facilities, their timing accuracies have
shown a sharp decreasing trend \citep{Liu:2022mcd}, which is unfavorable for
multi-messenger early-warning detections. Furthermore, their detection results
do not play a major part in our simulations for both two GW detectors.
Therefore, we only use the early-warning BNS merger samples in category (b).
Those events not only make up the majority of the total BNS mergers but also
yield the best and most stable estimation results.

Following \citet{Liu:2022mcd}, we use the Fisher  information matrix  to
estimate 9 system parameters, collectively denoted as
$\boldsymbol{\Xi}=\left\{\mathcal{M}, \eta, t_{\mathrm{c}}, \phi_{\mathrm{c}},
D_{\mathrm{L}}, \bar{\theta}_{\mathrm{S}}, \bar{\phi}_{\mathrm{S}},
\bar{\theta}_{\mathrm{L}}, \bar{\phi}_{\mathrm{L}}\right\}$, to obtain the
parameter precisions of each event. In $\boldsymbol{\Xi}$, $\eta \equiv M_{1}
M_{2} /\left(M_{1}+M_{2}\right)^{2}$ is the symmetric mass ratio; $\mathcal{M}
\equiv\left(M_{1}+M_{2}\right) \eta^{3 / 5}$ is the  chirp mass;
$t_{\mathrm{c}}$ and $\phi_{\mathrm{c}}$ are the time and orbital phase at
coalescence, respectively; and $\left\{\bar{\theta}_{\mathrm{S}},
\bar{\phi}_{\mathrm{S}}, \bar{\theta}_{\mathrm{L}},
\bar{\phi}_{\mathrm{L}}\right\}$ are the source direction and angular momentum
direction in the Solar system barycentric frame \citep[see Fig.~1 in][]{Liu:2020nwz}. 
From $\left\{\bar{\theta}_{\mathrm{S}},
\bar{\phi}_{\mathrm{S}}, \bar{\theta}_{\mathrm{L}},
\bar{\phi}_{\mathrm{L}}\right\}$ we can obtain the viewing angle
$\theta_{\mathrm{obs}}$  in the next section. The signal-to-noise
ratio (SNR) threshold is set to be 8 in this work. Our attention focuses on the estimation of
the accuracy of angular resolution $\Delta \Omega$ and time of merger $\Delta
t_{\mathrm{c}}$. For the Fisher matrix in the frequency domain, we set the
integration limit to be $f_{\text {in}}$ and $f_{\text {out}}$. As a BNS is to
merge in time $t_{\mathrm{c}_{0}}$, we have $f_{\text
{in}}=\left(t_{\mathrm{c}_{0}} / 5\right)^{-3 / 8} \mathcal{M}^{-5 / 8} / 8 \pi$
and $f_{\text {out}}=\left(t_{\mathrm{e}} / 5\right)^{-3 / 8} \mathcal{M}^{-5 /
8} / 8 \pi$ with $t_{\text {e}}$ representing the early-warning time before the
merger. The early-warning time $t_{\text {e}}$ is remarkably significant to the
multi-messenger early-warning detections of BNS mergers. We mainly present the
results with $t_{\mathrm{e}} = 1\,\mathrm{d}$ in detail, and analyses with a
larger $t_{\mathrm{e}}$ will be briefly mentioned.

\section{ EM Counterpart Detection } 
\label{ sec:EM Counterpart Detection }

Based on the GW early-warning populations, we explore the multi-messenger
observations with the multi-band EM follow-ups. Following \citet{Song:2019ddw}
and \citet{Yu:2021nvx}, in Sec.~\ref{ sec:sGRB Detection } we first briefly
introduce the method of sGRB detections, which are expected to occur shortly
after the BNS mergers. Compared to the kilonovae, there have already been many
cosmological sGRB afterglows observed ranging from radio to X-rays to date
\citep{Zhang:2018ads}. Thus in Sec.~\ref{ sec:Kilonova Detection } we focus on
the early-warning kilonova detections and report our results with detailed
analyses.

\subsection{ sGRB Detection }
\label{ sec:sGRB Detection }

\begin{table}
    \renewcommand\arraystretch{2}
    \centering
    \caption{Parameters of GRB\,170817A afterglow using Gaussian jet model
    \citep{Ryan:2019fhz}.}
    \begin{threeparttable}
    \setlength{\tabcolsep}{9.0mm}{\begin{tabular}{cc}
    \toprule
    \toprule
    Parameter    & Value \\
    \midrule 
    $\theta_{\mathrm{c}}/\mathrm{rad}$        & $0.066_{-0.018}^{+0.018}$\\
    $\theta_{\mathrm{w}}/\mathrm{rad}$        & $0.47_{-0.19}^{+0.26}$\\
    $\log _{10} (E_{0}/\mathrm{erg})$        & $52.96_{-0.72}^{+0.97}$\\
    $\log _{10} (n_{0}/\mathrm{cm}^{-3})$        & $-2.7_{-1.0}^{+1.0}$\\
    $p$        & $2.1675_{-0.0075}^{+0.0063}$\\
    $\log _{10} \varepsilon_{\mathrm{e}}$        & $-1.4_{-1.1}^{+0.7}$\\
    $\log _{10} \varepsilon_{\mathrm{B}}$        & $-4.0_{-0.7}^{+1.1}$\vspace{0.3em}\\ 
    \bottomrule
    \end{tabular}}
    \begin{tablenotes}
      \item \textbf{Note.} $n_{0}$ is the circumburst density; ${p}$ is the
      electron energy index; $\varepsilon_{\mathrm{e}}$ and
      $\varepsilon_{\mathrm{B}}$ are the fractions of shock energy carried by
      electrons and magnetic fields, respectively. The given posterior values
      are the medians, along with their 16\%, and 84\% quantiles.
    \end{tablenotes}
    \end{threeparttable}
    \label{ tab:Posterior Distribution }
\end{table}

The observational fact that GRB\,170817A is abnormally less energetic than
typical sGRBs \citep{Goldstein:2017mmi, 2018NatCo...9..447Z} with the slowly
rising multi-band light curves \citep{Troja:2017nqp, Troja:2018ruz,
Lazzati:2017zsj, Lyman:2018qjg, 2018Natur.554..207M} suggested that the scenario
of BNS merger is most likely to be an off-axis configuration
\citep{2002MNRAS.332..945R, 2002ApJ...571..876Z}. Among a variety of jet energy
profiles, we adopt the Gaussian structured jet model
\citep{2002ApJ...571..876Z},
\begin{equation}
E(\theta)=E_{0} \exp \left(-\frac{\theta^{2}_{\mathrm{obs}}}{2
\theta_{\mathrm{c}}^{2}}\right)\,, \quad\quad \theta_{\mathrm{obs}} \leqslant
\theta_{\mathrm{w}} \,,
\label{ eq:Gaussian-shaped jet profile }
\end{equation}
where $E_0$ is the on-axis equivalent isotropic energy, $\theta_{\mathrm{obs}}$
is the polar viewing angle, $\theta_{\mathrm{c}}$ is the characteristic core
angle, and $\theta_{\mathrm{w}}$ is the truncation angle of the jet, which means
that the energy would be initially zero for $\theta_{\mathrm{obs}} >
\theta_{\mathrm{w}}$.  \add{This model is favored by former analyses
\citep{Lazzati:2017zsj, 2018Natur.554..207M, Troja:2018ruz, Troja:2018uns,
Troja:2020pzf, Xie:2018vya, Lamb:2018qfn}.}  \citet{Ryan:2019fhz} has given the
constraints on the jet and afterglow parameters for GRB\,170817A, which are
listed in Table~\ref{ tab:Posterior Distribution }. Throughout this paper, all
the parameters in Table~\ref{ tab:Posterior Distribution } are fixed to their
median values for simplification.

\begin{table*}
    \renewcommand\arraystretch{1.8}
    \centering
    \caption{Summary of technical information for each proposed survey from
    \citet{Zhu:2021ram, Zhu:2020ffa} and CSST-HOD, assuming a 300-s exposure
    time for the search limiting magnitude $m^*$.} 
    \setlength{\tabcolsep}{0.87cm}{\begin{tabular}{c ccc c c}
    \toprule
    \toprule
    Telescope        & \multicolumn{3}{c}{$m^*$}
                     & FoV/$\mathrm{deg}^2$    
                     & Reference\vspace{-0.8em}\\
                     & $\it{g}$ & $\it{r}$ & $\it{i}$\\
    \midrule
    WFST       & 24.18        & 23.95        & 23.33        & 6.55         
               & \citet{2018AcASn..59...22S}\\
    LSST       & 26.15        & 25.70        & 25.79        & 9.6        
               & \citet{LSSTScience:2009jmu} \\
    CSST       & 26.3         & 26.0         & 25.9         & 1.1         
               & \citet{Gong:2019yxt} \\
    CSST-HOD   & 27           & 27           & 27           & --
               & --\vspace{0.3em}\\ 
    \bottomrule
    \end{tabular}}
    \label{ tab:Survey projects }
\end{table*}

Following \citet{Yu:2021nvx}, we assume that every early-warning BNS merger
event in our simulation is associated with an sGRB, whose jet profile is broadly
similar to that of GRB\,170817A in Eq.~(\ref{ eq:Gaussian-shaped jet profile }).
Assuming that the burst duration $T_{90} \sim 2$\,$\mathrm{s}$
\citep{LIGOScientific:2017ync} and the spectrum is flat with time during
$T_{90}$, the $\gamma$-band flux for each BNS merger is,
\begin{equation}
F_{\gamma}=\frac{E_{0} \eta_{\gamma}}{4 \pi D_{\mathrm{L}}^{2}T_{90}} \exp
\left(-\frac{\theta_{\mathrm{obs}}^{2}}{2 \theta_{\mathrm{c}}^{2}}\right)\,,
\quad \ \theta \leqslant \theta_{\mathrm{w}} \,,
\label{ eq:flux }
\end{equation}
where $\eta_{\gamma}$ is the radiative efficiency and we use $\eta_{\gamma} \sim
0.1$ for the bolometric energy flux in the $1$--$10^{4}\,\mathrm{keV}$ band
\citep{Yu:2021nvx}. For all BNS-associated sGRBs, we assume that the
$\gamma$-ray spectrum is described by the Band function with photon indices
$\alpha = -1$ and $\beta = -2.3$ respectively below and above the peak energy
$E_{\mathrm{p}}$ in the $\nu F_{\nu}$ spectra \citep{Preece:1999fv}. Note that
$E_{\mathrm{p}}$ changes with $\theta_{\mathrm{obs}}$ via the relationship
proposed by \citet{Ioka:2019jlj} to make the GRB\,170817A observation consistent
with previous sGRB data via, 
\begin{equation}
E_{\mathrm{p}}(\theta_{\mathrm{obs}})=E_{\mathrm{p}, 0}
\left(1+\theta_{\mathrm{obs}} / \theta_{\mathrm{c}}\right)^{-0.4}\,,
\label{ eq:IN relationship}
\end{equation}
where $E_{\mathrm{p}, 0}$ can be calculated by the Yonetoku relation
\citep{Yonetoku:2003gi} with the central luminosity of the Gaussian jet. 

Adopting the same setting of four $\gamma$-ray detectors as in
\citet{Yu:2021nvx}, the sensitivity for the Fermi Gamma-ray Burst Monitor
(Fermi-GBM) is $\sim 2 \times 10^{-7}\,{\rm erg\,s}^{-1}\,\mathrm{cm}^{-2}$ in
the $50$--$300\,\mathrm{keV}$ band \citep{Meegan:2009qu}; the sensitivity for
the Gravitational wave high-energy Electromagnetic Counterpart All-sky Monitor
(GECAM) is $\sim 1 \times 10^{-7}\,{\rm erg\,s}^{-1}\,\mathrm{cm}^{-2}$ in the
$50$--$300\,\mathrm{keV}$ band \citep{Zhang:2018hzq}; \add{the sensitivity for
the Swift Burst Alert Telescope (BAT) and the Space Variable Objects Monitor
(SVOM)-ECLAIRS is $\sim 1.2 \times
10^{-8}\,\operatorname{erg\,s}^{-1}\,\mathrm{cm}^{-2}$ in the
$15$--$150\,\mathrm{keV}$ band \citep{SwiftScience:2004ykd, Lien:2013qja,
Gotz:2015mha}.} Based on the above assumptions, we can obtain the effective
sensitivity limit for different $\gamma$-ray detectors in the
$1$--$10^{4}\,\mathrm{keV}$ band. By comparing $F_{\gamma}$ with the effective
sensitivity limit, we can then assess whether the simulated sGRB could be
detected by a specific $\gamma$-ray detector. As shown in Fig.~2 of
\citet{Song:2019ddw}, the effective sensitivity for Swift-BAT and SVOM-ECLAIRS
(denoted as ``SBSE'' hereafter) performs the best in the
$1$--$10^{4}\,\mathrm{keV}$ band, while Fermi-GBM (denoted as ``FG'' hereafter)
has the worst performance. Although these $\gamma$-ray detectors might have no
overlapping observational time with our space-borne decihertz GW observatories
(B-DEC and DO-OPT), our results can still apply to the similar detectors at that
time. Furthermore, for each detector mentioned above, we also have considered
hypothetically optimized devices (HODs). We artificially set their sensitivities
to be one (HOD-1) and two (HOD-2) orders of magnitude better than the current
ones. The results on the sGRB detections are given in Sec.~\ref{ sec:Results }.

\subsection{ Kilonova Detection }
\label{ sec:Kilonova Detection }

\begin{figure}
    \centering
    \includegraphics[width=8.5cm]{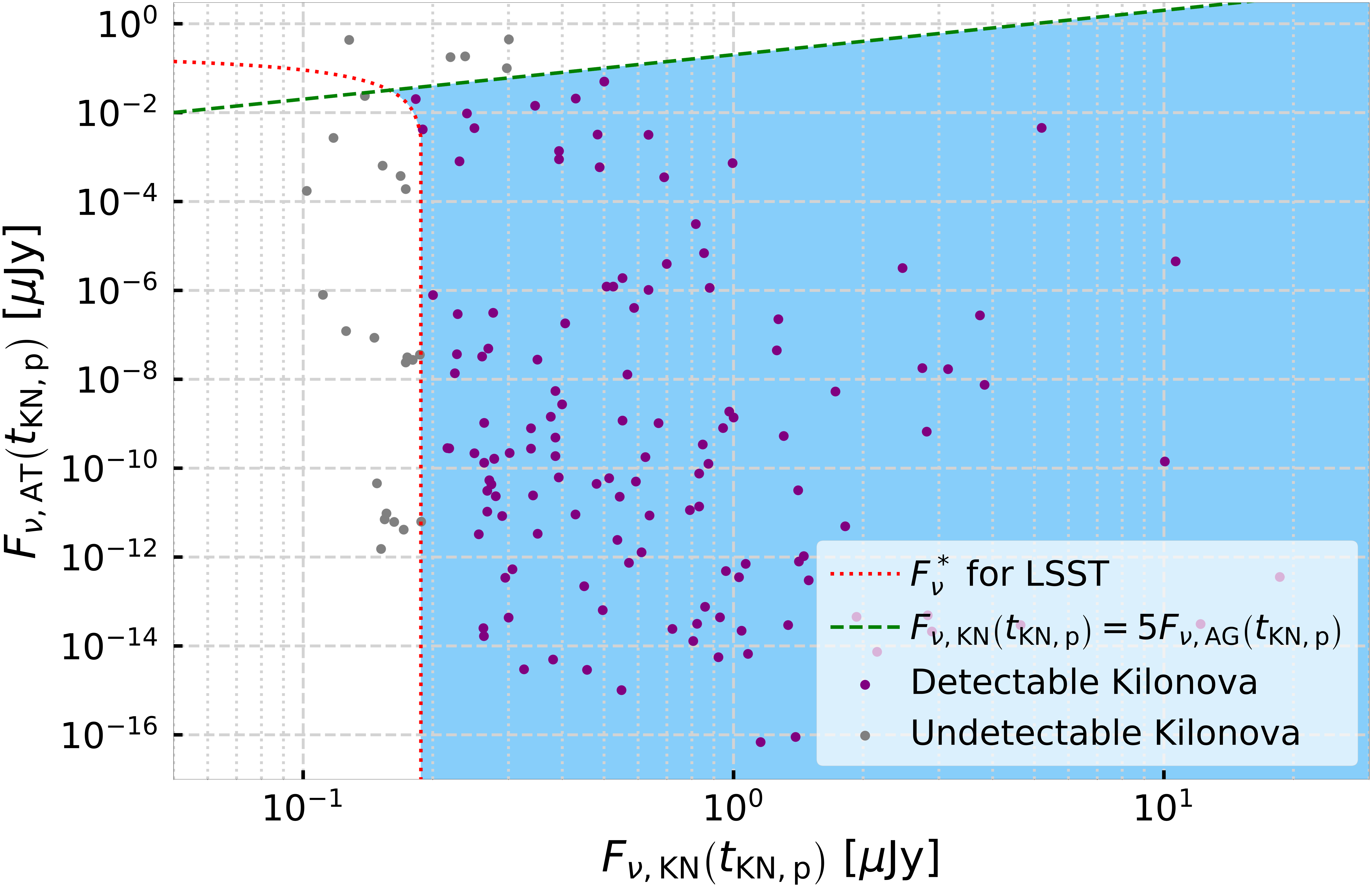}
    \caption{Kilonova detection parameter space for $r$-band of LSST. The shaded
    blue region delimits the detectable $F_{\nu, \mathrm{KN}}$-$F_{\nu,
    \mathrm{AG}}$ space  according to the two criteria in the main text. The GW
    early-warning kilonovae  in this figure are obtained from B-DEC with
    $2\,\mathrm{yr}<t_{\mathrm{c}_{0}} \leq 3\,\mathrm{yr}$, assuming the LN
    population model and $t_\mathrm{e} = 1\,\mathrm{d}$.} \label{ fig:kilonova
    criteria }
\end{figure}

\citet{Li:1998bw} predicted thermal, supernova-like transients with a duration 
of a day to a week after BNS mergers. Since then the kilonova provides both a
candidate EM counterpart to the GW-triggered BNS mergers and a probe of the
astrophysical origin of $r$-process elements \citep{Metzger:2010sy,
Metzger:2016pju}. However, it can be difficult to detect the cosmological
kilonovae not associated with beamed sGRBs and afterglows due to their
relatively low luminosity and fast-evolving nature. In addition to AT\,2017gfo
with relatively complete observations of kilonova properties, \add{there is no
kilonova candidate for the third observing run (O3)  of LIGO/Virgo/KAGRA
\citep{Kasliwal:2020wmy, Becerra:2021lyq, deJaeger:2021tcq, Dichiara:2021vjy,
Mohite:2021pfn}.} Therefore, following \citet{Zhu:2021zmy}, we  simply assume
that all kilonovae are AT\,2017gfo-like in view of the scarce data to
date.\footnote{\add{Note that there are many theoretical works suggesting that
BNS kilonovae should be diverse, depending on the mass ratio of the binary and
the nature of the merger remnant \citep{Kasen:2013xka, Kasen:2017sxr,
Kawaguchi:2019nju, Kawaguchi:2020vbf, Korobkin:2020spe}. Also, some recent
observational data have shown evidence for the deviations from AT 2017gfo
\citep{Tanvir:2013pia, Troja:2019ccb, OConnor:2020qmy}. In view of this, we
caution that some results in this paper can be affected by the diversity of the
kilonovae. A future dedicated investigation is required to address this aspect
fully.}} Adopting the model generated by the \texttt{POSSIS} package
\citep{Bulla:2019muo, Coughlin:2020ozl}, which considered a polar-dominated
lanthanide-free (LF) component and an equatorial-dominated lanthanide-rich (LR)
component with different opacities, we can obtain the multi-band kilonova data
with a total ejecta mass $M_{\mathrm{ej}}=0.04\,M_{\odot}$ and a half-opening
angle of the LR component $\Phi=60^{\circ}$ \citep[for example, see Fig.~2
in][]{Zhu:2021zmy}.

\begin{table*}
    \renewcommand\arraystretch{1.8}
    \centering
    \caption{Yearly joint GW-sGRB detection numbers and percentages observed by
    different $\gamma$-ray missions and two decihertz GW detectors for different
    early-warning BNS merger populations. We assume an early-warning time
    $t_\mathrm{e} = 1\,\mathrm{d}$. We list the total number of early-warning
    BNS mergers per year in the parentheses below B-DEC and DO-OPT, which are
    used in the percentage calculations. In each cell we list
    the result with the current design for each $\gamma$-ray detector as the
    first entry, and present results for HOD-1 and HOD-2 as the second and the
    third entries. These values are separated by two slashes.}  
    \setlength{\tabcolsep}{0.55cm}{\begin{tabular}{c c c c c}
    \toprule
    \toprule
    Population  & Detector   & FG    & GECAM    & SBSE\\
    
    \midrule
    
    \multirow{4}{15pt}{\shortstack{LN}}
    & \multirow{2}{30pt}{\shortstack{\\B-DEC\\ \\(166)}} 
    & 5 / 12 / 19    & 6 / 15 / 20    & 10 / 19 / 25\vspace{-0.3em}\\
    &            
    & (3.0\,\% / 7.2\,\% / 11\,\%)   & (3.6\,\% / 9.0\,\% / 12\,\%) 
    & (6.0\,\% / 11\,\% / 15\,\%)\\
    & \multirow{2}{44pt}{\shortstack{\\DO-OPT\\($1.3 \times 10^{4}$)}}
    & 10 / 158 / 465    & 28 / 243 / 554    & 120 / 428 / 736\vspace{-0.3em}\\
    &            
    & ($<$0.1\,\% / 1.3\,\% / 3.7\,\%)   & (0.2\,\% / 1.9\,\% / 4.4\,\%)
    & (1.0\,\% / 3.4\,\% / 5.9\,\%)
    \vspace{0.5em}
    \\
    
    \midrule
    
    \multirow{4}{30pt}{\shortstack{SFR14}}
    & \multirow{2}{30pt}{\shortstack{\\B-DEC\\ \\(127)}} 
    & 3 / 7 / 14    & 4 / 9 / 14    & 6 / 12 / 16\vspace{-0.3em}\\
    &            
    & (2.4\,\% / 5.5\,\% / 11\,\%)   & (3.1\,\% / 7.1\,\% / 11\,\%)
    & (4.7\,\% / 9.4\,\% / 13\,\%)\\
    & \multirow{2}{44pt}{\shortstack{\\DO-OPT\\($1.5 \times 10^{4}$)}}
    & 10 / 160 / 678    & 22 / 296 / 842    & 115 / 622 / $1.2 \times 10^{3}$\vspace{-0.3em}\\
    &            
    & ($<$0.1\,\% / 1.1\,\% / 4.7\,\%)   & (0.2\,\% / 2.0\,\% / 5.8\,\%)
    & (0.8\,\% / 4.3\,\% / 8.1\,\%)\vspace{0.5em}\\
    
    \midrule
    
    \multirow{4}{42pt}{\shortstack{Stan.High}}
    & \multirow{2}{30pt}{\shortstack{\\B-DEC\\ \\(122)}} 
    & 5 / 8 / 12    & 6 / 10 / 13    & 7 / 12 / 14\vspace{-0.3em}\\
    &            
    & (4.1\,\% / 6.6\,\% / 9.8\,\%)   & (4.9\,\% / 8.2\,\% / 11\,\%)
    & (5.7\,\% / 9.8\,\% / 11\,\%)\\
    & \multirow{2}{44pt}{\shortstack{\\DO-OPT\\($9.7 \times 10^{3}$)}}
    & 6 / 118 / 390    & 18 / 198 / 475    & 84 / 365 / 661\vspace{-0.3em}\\
    &            
    & ($<$0.1\,\% / 1.2\,\% / 4.0\,\%)   & (0.2\,\% / 2.0\,\% / 4.9\,\%)
    & (0.9\,\% / 3.8\,\% / 6.8\,\%)\vspace{0.5em}\\
    
    \midrule
    
    \multirow{4}{38pt}{\shortstack{Oce.High}}
    & \multirow{2}{30pt}{\shortstack{\\B-DEC\\ \\(332)}} 
    & 9 / 16 / 26    & 10 / 20 / 29    & 14 / 24 / 35\vspace{-0.3em}\\
    &            
    & (2.7\,\% / 4.8\,\% / 7.8\,\%)   & (3.0\,\% / 6.0\,\% / 8.7\,\%)
    & (4.2\,\% / 7.2\,\% / 11\,\%)\\
    & \multirow{2}{44pt}{\shortstack{\\DO-OPT\\($3.3 \times 10^{4}$)}}
    & 17 / 342 / $1.5 \times 10^{3}$    & 44 / 642 / $1.8 \times 10^{3}$    & 234 / $1.4 \times 10^{3}$ / $2.6 \times 10^{3}$\vspace{-0.3em}\\
    &            
    & ($<$0.1\,\% / 1.0\,\% / 4.5\,\%)   & (0.1\,\% / 2.0\,\% / 5.6\,\%)
    & (0.7\,\% / 4.1\,\% / 7.8\,\%)\vspace{0.3em}\\ 
    \bottomrule
    \end{tabular}}
    \label{ tab:sRGB detection rates }
\end{table*}

The presence of relatively bright jet afterglows may affect observations of
kilonovae for on-axis or near-axis observers. Therefore, we
define a detectable kilonova event with two criteria \citep{Zhu:2021ram},
\begin{enumerate}[(i)] 
  \item $F_{\nu, \mathrm{KN}}\left(t_{\mathrm{KN}, \mathrm{p}}\right)+F_{\nu,
  \mathrm{AG}}\left(t_{\mathrm{KN}, \mathrm{p}}\right)>F_{\nu}^*$, and
  \item $F_{\nu, \mathrm{KN}}\left(t_{\mathrm{KN}, \mathrm{p}}\right)>5 F_{\nu,
  \mathrm{AG}}\left(t_{\mathrm{KN}, \mathrm{p}}\right)$,
\end{enumerate}
where $t_{\mathrm{KN}, \mathrm{p}}$ is the peak time of the kilonova; $F_{\nu,
\mathrm{KN}}$ and $F_{\nu, \mathrm{AG}}$ are the peak kilonova flux and the
afterglow flux, respectively; $F_{\nu}^*$ is the effective limiting flux for
different survey telescopes. Three proposed survey projects with three most
common filters ($\it{gri}$) have been considered in our work, which are the Wide
Field Survey Telescope \citep[WFST;][]{2018AcASn..59...22S}, the Large Synoptic
Survey Telescope \citep[LSST, newly named as the Vera Rubin
Observatory;][]{LSSTScience:2009jmu}, and the Chinese Space Station Telescope
\citep[CSST;][]{Gong:2019yxt}. Some technical parameters for these proposed
survey telescopes are given in Table~\ref{ tab:Survey projects }.  We could
obtain $F_{\nu}^* \simeq 3631\,\mathrm{Jy} \times 10^{-m^* / 2.5}$ in each band
when the search limiting magnitude $m^*$ is given. Note that a 300-s exposure
time is adopted throughout this paper and we again consider a HOD for CSST
(CSST-HOD) with one magnitude better sensitivity than CSST in each band
(see Table~\ref{ tab:Survey projects }).  

Taking the LN population model with $t_\mathrm{e} = 1\,\mathrm{d}$ for B-DEC as
an example, we plot in Fig.~\ref{ fig:kilonova criteria } the selection
functions corresonding to the criteria~(i) and (ii) above. The red dotted line
corresponds to the $\it{r}$-band effective limiting flux $F_{\nu}^*$ of LSST,
and the green dashed line indicates the boundary of criterion~(ii). Therefore,
the shaded blue region in Fig.~\ref{ fig:kilonova criteria } delimits the
detectable $F_{\nu, \mathrm{KN}}$-$F_{\nu, \mathrm{AG}}$ parameter space of
kilonovae for LSST.

For all the valuable early-warning joint GW-kilonova observations, we further
classify them into two groups, ``$GK$'' and ``$K$'', based on the observation of
sGRB and kilonova, according to
\begin{enumerate}[(I)]
  \item $GK$: both sGRB and kilonova can be detected;
  \item $K$: kilonova can be detected while sGRB cannot.
\end{enumerate}
Here we mention that, observing $\it{K}$ samples is indeed significant, but
without the synergy with sGRB as $\it{GK}$ samples,  sometimes it can be 
difficult to distinguish kilonovae from other rapid-evolving transients, for
example, some supernova events \citep{Mazzali:2008zd, Prentice:2018qxn,
Perley:2018oky, McBrien:2019wfv, Chen:2019bjf}. Therefore, more attention should
be paid to the $\it{GK}$ samples in future work. In this work we use
\texttt{afterglowpy}, an open-source Python package \citep{Ryan:2019fhz}, to
model the light curves of multi-wavelength afterglow and obtain the $F_{\nu,
\mathrm{AG}}$ value with parameters listed in Table~\ref{ tab:Posterior
Distribution }. Note that for $\it{GK}$ and $\it{K}$ samples, the definition of
sGRB detection is based on the detection result of SBSE, except for CSST-HOD.
We consider the SBSE-HOD-1 and CSST-HOD to achieve more cooperative observations
during the same period. 

\section{ Results } 
\label{ sec:Results }

\begin{table*}
    \renewcommand\arraystretch{1.8}
    \centering
    \caption{Yearly joint GW-kilonova detection numbers and percentages observed
    by different survey missions and two decihertz GW detectors for our
    early-warning BNS mergers. We have used  the ``LN'' population model with
    $t_\mathrm{e} = 1\,\mathrm{d}$ as an example. Three values in each cell for
    $\it{GK}$ and $\it{K}$ samples represent the results with $\it{g/r/i}$
    filters. CSST-HOD has approximately one magnitude better
    sensitivity than CSST (see Table~\ref{ tab:Survey projects } in  Sec.~\ref{
    sec:Kilonova Detection }). We have calculated the total number of BNS merger
    detections, as  listed in Table~\ref{ tab:sRGB detection rates } for B-DEC
    (166) and DO-OPT ($1.3 \times 10^{4}$). For WFST and LSST, the $1/4$
    discount due to the sky coverage $\Omega_{\mathrm{cov}}$ and the Earth's
    rotation is not considered (see Sec.~\ref{ sec:Detection Rates }).}
    \setlength{\tabcolsep}{0.17cm}{\begin{tabular}{c c c c c c}
    \toprule
    \toprule
    Type   & Detector    & WFST    & LSST    & CSST    
                     & CSST-HOD\\
    \midrule
    
    \multirow{4}{15pt}{\shortstack{$GK$}}
    & \multirow{2}{30pt}{\shortstack{B-DEC}} 
    & $<$\,1 / 2 / 1    &  3 / 4 / 5    & 3 / 4 / 5    & 11 / 13 / 14\vspace{-0.5em}\\
    &          
    & ($<$0.6\,\% / 1.2\,\% / 0.6\,\%)            & (1.8\,\% / 2.4\,\% / 3.0\,\%)          
    & (1.8\,\% / 2.4\,\% / 3.0\,\%)            & (6.6\,\% / 7.8\,\% / 8.4\,\%)\\
    & \multirow{2}{37pt}{\shortstack{DO-OPT}}
    & $<$\,1 / 1 / 1    & 2 / 3 / 5    & 2 / 4 /  5    
    & 12 / 31 / 43\vspace{-0.5em}\\
    &            
    & ($<$0.1\,\% / $<$0.1\,\% / $<$0.1\,\%)           & ($<$0.1\,\% / $<$0.1\,\% / $<$0.1\,\%)   
    & ($<$0.1\,\% / $<$0.1\,\% / $<$0.1\,\%)          & (0.1\,\% / 0.2\,\% / 0.3\,\%)\vspace{0.5em}\\
    
    \midrule
    
    \multirow{4}{9pt}{\shortstack{$K$}}
    & \multirow{2}{30pt}{\shortstack{B-DEC}} 
    & 23 / 35 / 30    & 111 / 138 / 153    & 123 / 150 / 155 
    & 143 / 147 / 147\vspace{-0.5em}\\
    &            
    & (14\,\% / 21\,\% / 18\,\%)         & (67\,\% / 83\,\% / 92\,\%)   
    & (74\,\% / 90\,\% / 93\,\%)         & (86\,\% / 89\,\% / 89\,\%)\\
    & \multirow{2}{37pt}{\shortstack{DO-OPT}}
    & 16 / 29 / 22    & 181 / 253 / 485    & 219 / 361 / 544
    & 517 / $1.2 \times 10^{3}$ / $1.7 \times 10^{3}$\vspace{-0.5em}\\
    &            
    & (0.1\,\% / 0.2\,\% / 0.2\,\%)           & (1.4\,\% / 2.0\,\% / 3.9\,\%)   
    & (1.7\,\% / 2.9\,\% / 4.3\,\%)          & (4.1\,\% / 9.4\,\% / 14\,\%)\vspace{0.5em}\\
    \bottomrule
    \end{tabular}}
    \label{ tab:kilonova detection rates }
\end{table*}

Now we give detailed results for the multi-messenger early-warning detections of
BNS mergers. Given that the  merger events in our consideration from
category~(b) with $1\,\mathrm{yr}<t_{\mathrm{c}_{0}} \leq 4\,\mathrm{yr}$ are in
general distributed uniformly in time \citep[see e.g., Fig.~3 and Fig.~7
in][]{Liu:2022mcd}, we divide the total GW early-warning events by 3 and present
the yearly sGRB and kilonova detection rates in Sec.~\ref{ sec:Detection Rates
}. In Sec.~\ref{ sec:Parameter Distributions }, we plot the characteristic
distributions of the yearly multi-messenger events. Note that for all the
scatter plots, we choose the early-warning BNS mergers with
$2\,\mathrm{yr}<t_{\mathrm{c}_{0}} \leq 3\,\mathrm{yr}$.

\subsection{ Detection Rates }
\label{ sec:Detection Rates }

As shown in Table~\ref{ tab:sRGB detection rates } we summarize all our
simulated joint GW-sGRB detections by different $\gamma$-ray missions and two
decihertz GW detectors. These results are for different BNS merger populations
with an early warning time $t_\mathrm{e} = 1\,\mathrm{d}$. With a
GRB\,170817A-like Gaussian jet structure, we can see that fewer than 10\% of the
early-warning BNS mergers for B-DEC would have a $\gamma$-band flux higher than
the threshold of the $\gamma$-ray detectors with current designs. However, with
the improved sensitivity by one (two) order(s) of magnitude for each
$\gamma$-ray detector---namely, HOD-1 (HOD-2) in our notations---we find B-DEC
could approximately double (treble) the joint sGRB detection rates, as shown in
Table~\ref{ tab:sRGB detection rates }.  Given that DO-OPT has a lower noise
level \citep[see e.g., Fig.~2 in][]{Liu:2022mcd}, we find that it shows better
detection abilities, especially for more detections at higher redshift, as we
will elaborate later in Sec.~\ref{ sec:Parameter Distributions }. Furthermore,
with a larger $t_{\mathrm{e}}$, we have verified that the detection rates would
decrease due to a shorter stay in the decihertz  GW band. For example, all the
results could be halved for B-DEC with $t_{\mathrm{e}} = 14\,\mathrm{d}$. 

\begin{figure*}[ht]
    \centering
    \includegraphics[width=8.2cm]{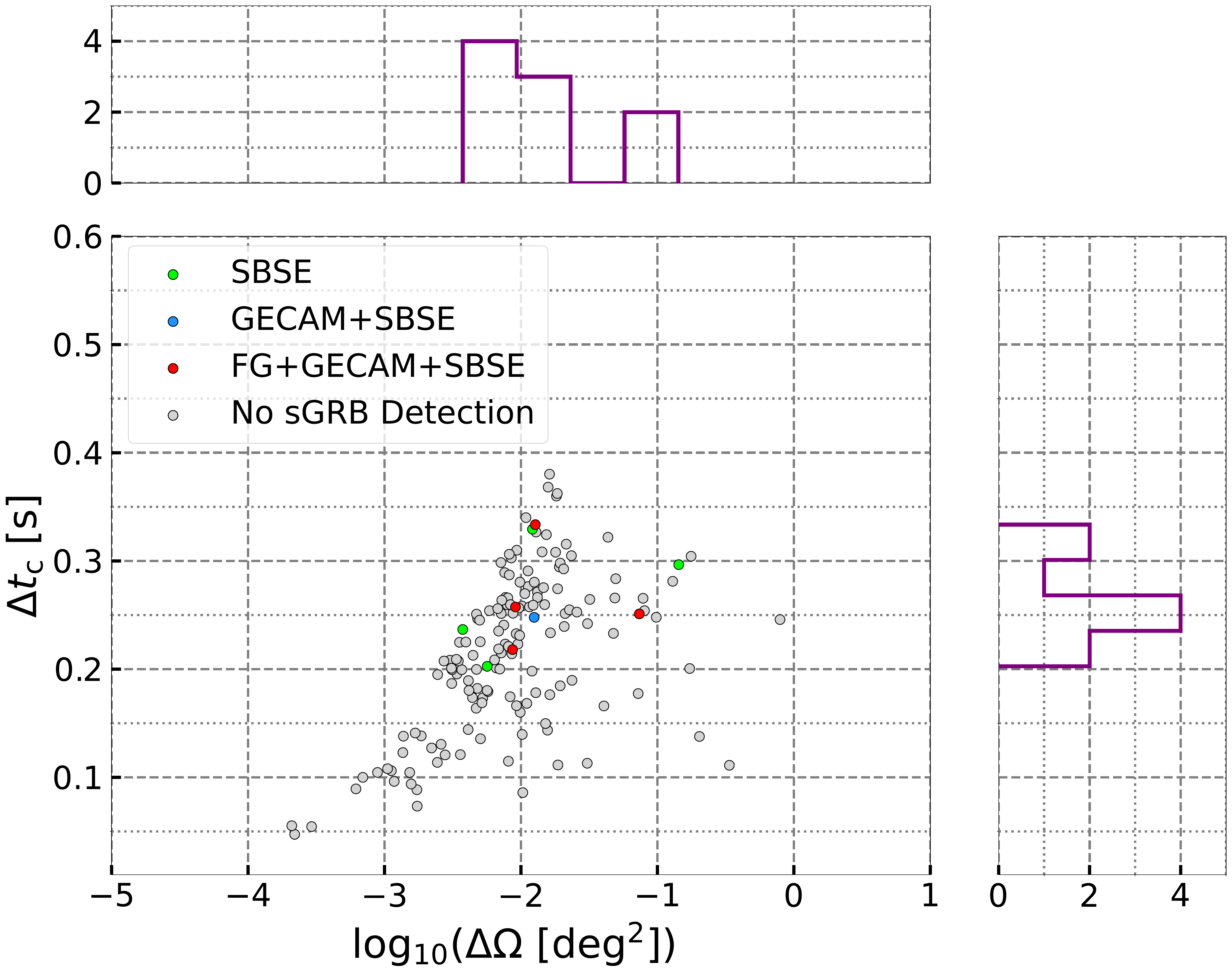}
    \hspace{2.5em}
    \includegraphics[width=8.2cm]{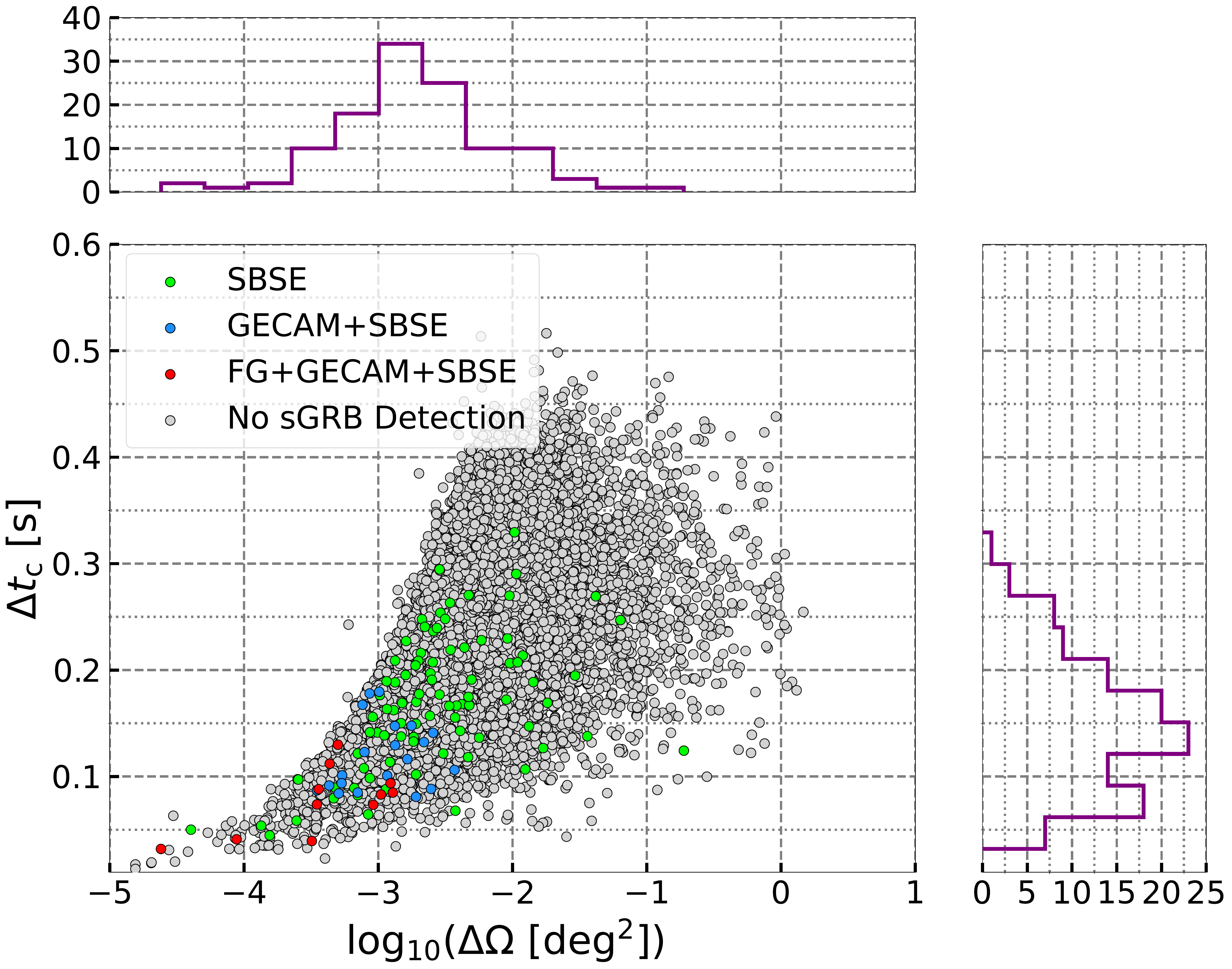}
    
    \vspace{1em}
    
    \includegraphics[width=8.2cm]{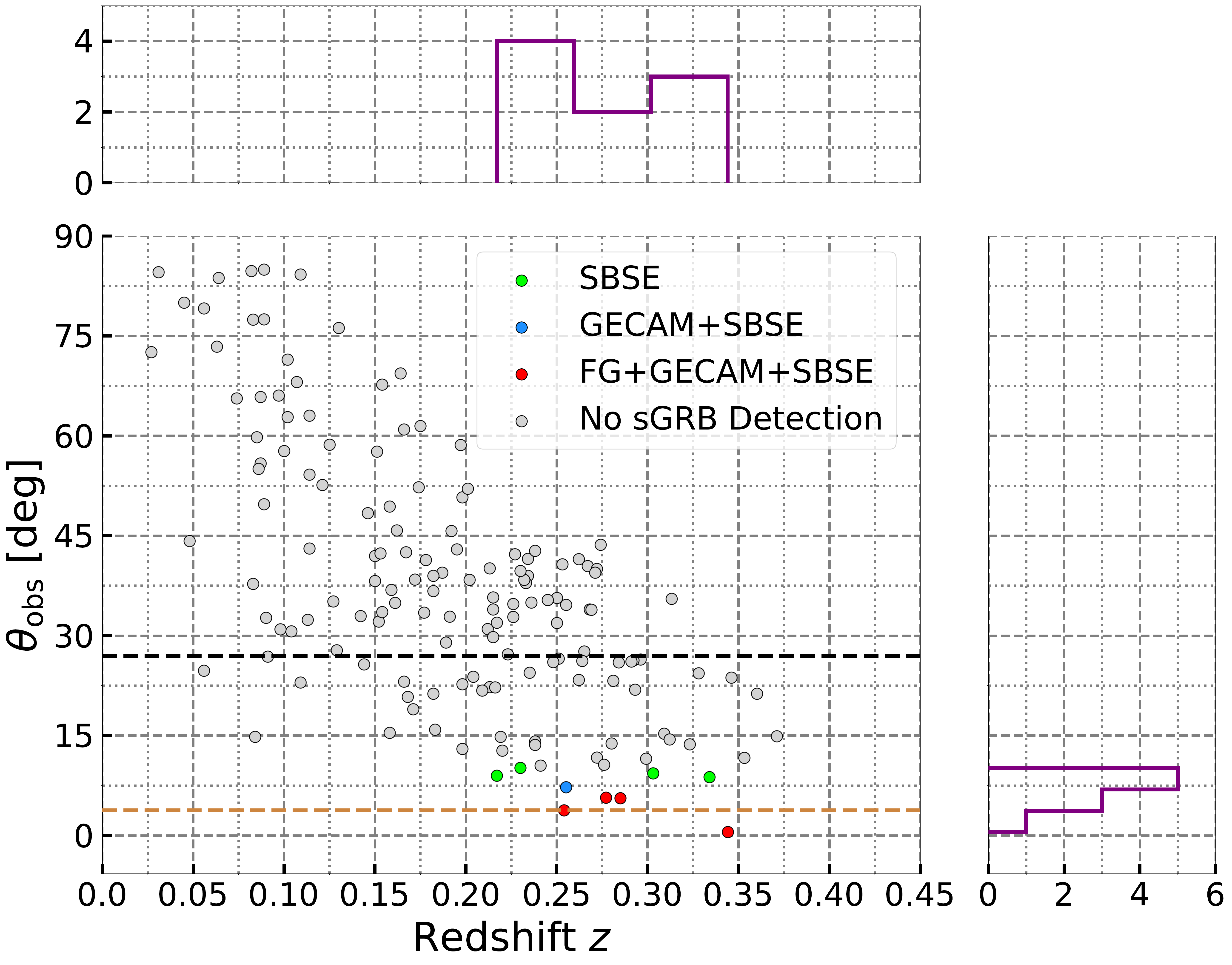}
    \hspace{2.5em}
    \includegraphics[width=8.2cm]{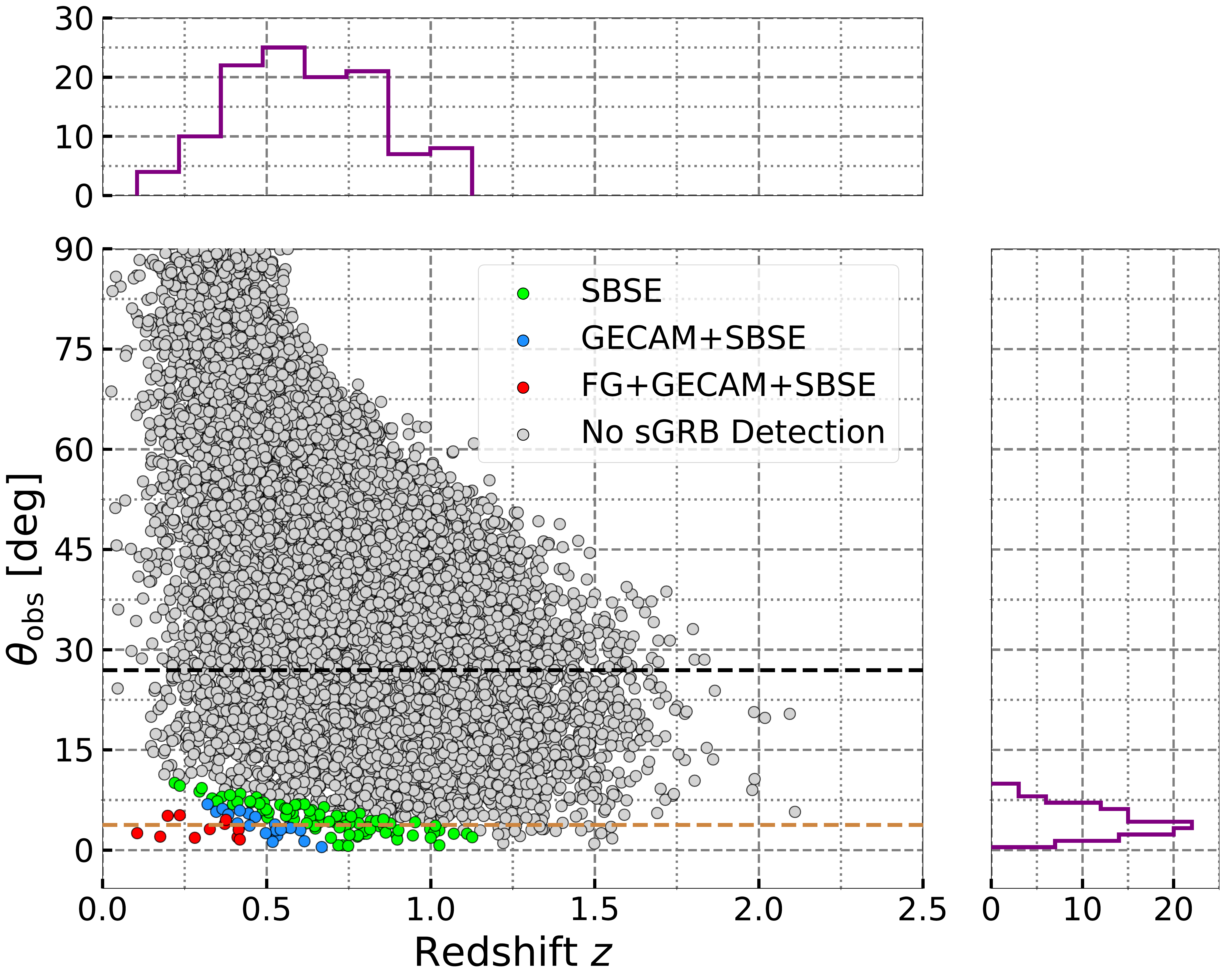}
    \caption{The $\Delta \Omega$-$\Delta t_{c}$ (top panels) and
    $z$-$\theta_{\mathrm{obs}}$ (bottom panels) distributions of our yearly
    early-warning events in the LN model with $t_\mathrm{e} = 1\,\mathrm{d}$ as
    an example. Left panels are for B-DEC while right panels are for DO-OPT. The
    dashed horizontal black  and brown lines in bottom panels correspond to the
    fixed values of $\theta_{\mathrm{w}}$ and $\theta_{\mathrm{c}}$,
    respectively (see Sec.~\ref{ sec:sGRB Detection } and Table~\ref{
    tab:Posterior Distribution }). We plot purple histograms for all the events
    with joint sGRB detections (i.e., including the green, blue and red
    circles). Gray circles represent the mergers without sGRB joint detections.
    } \label{ fig:sGRB distributions }
\end{figure*}

According to \citet{Song:2019ddw}, although FG can cover about three-quarters of
the whole sky and the FoV for GECAM is about $4\pi$, the FoV/$4\pi$ is $\sim
1/9$ for Swift-BAT and $\sim 1/5$ for SVOM-ECLAIRS, respectively. Considering
the short delay between the BNS merger and the sGRB detection, it inevitably
means that FoV could become an important discount factor to be reckoned with for
joint GW-sGRB detections. However, as we will see later in Sec.~\ref{
sec:Parameter Distributions } (e.g., Fig.~\ref{ fig:sGRB distributions }
therein), the sub-$\mathrm{deg}^2$ localization uncertainty region for decihertz
space-borne GW detectors could be totally covered by different $\gamma$-ray
detectors. Assuming that all the   $\gamma$-ray satellites are capable of
adjusting the pointing to the early-warning BNS mergers, there is no need to
consider the discounts from FoVs anymore like previous studies
\citep{Song:2019ddw, Yu:2021nvx}. We must highlight that with a sufficient
$t_\mathrm{e}$ prepared for a real sense of early-warning observations, such a
wait-for pattern will bring an excellent performance for the joint GW-sGRB
detection and future multi-messenger astronomy.

On the other hand, for the early-warning kilonova events, we list the yearly
detection results in Table~\ref{ tab:kilonova detection rates } using the LN
population model with $t_\mathrm{e} = 1\,\mathrm{d}$ as an example. Given a
specific combination of GW and EM detectors, we can compare the detection
results in different filters ($\it{gri}$), as separated by two slashes in the
table. Compared with the results in $\it{g/r}$-band, we find that the
$\it{i}$-band has seen the biggest increase in the proportion of the detection
rate from WFST to CSST-HOD, especially for DO-OPT. It can be understood as that
more early-warning events will be detected at higher redshift. Considering
different $m^*$ values for several optical survey missions in each band, we find
that the detection numbers of $\it{K}$ samples increase remarkably with the
improved sensitivity of the survey telescope, while for $\it{GK}$ samples only
CSST-HOD shows obvious improvement compared with the others. The discussion of
joint kilonova detections with a hypothetical CSST-HOD is to provide a rough
reference for the next-generation survey projects at the same time of decihertz
GW detectors.

Again, we assume that the space-borne survey telescopes (CSST and CSST-HOD) can
adjust the pointing to each early-warning BNS merger. However, for a specific
ground-based optical survey project, the maximum detectable sky coverage
$\Omega_{\mathrm{cov}}$ and the Earth's rotation must be taken into
consideration in the practical kilonova observations. For instance, if the
kilonovae are outside $\Omega_{\mathrm{cov}}$ or occur during the daytime, WFST
and LSST may not achieve the early-warning joint GW-kilonova observations, even
though we have a sufficient $t_\mathrm{e}$ for it. The
$\Omega_{\mathrm{cov}}/4\pi$ is $\sim 1/2$ for both WFST and LSST, and we
further assume a half-year maximum observation time to provide another $1/2$
discount for yearly events. Therefore, in a more realistic situation, we suggest that a factor of $1/4$ should be the total discount factor for the results of WFST and LSST in Table~\ref{ tab:kilonova detection rates }. Nevertheless, for the wait-for pattern discussed in this
work, we also point out that such a discount problem can be partially solved, e.g., with at least two survey telescopes on the different sides of the Earth. Moreover, the long duration of kilonovae (compared with sGRBs) also partially solves this problem. We hope that more studies can be carried out in the future to give more precise predictions with detailed analyses.

Compared with B-DEC in Table~\ref{ tab:kilonova detection rates }, we conclude
that DO-OPT performs better on the joint GW-kilonova detections as a whole. This
is because that DO-OPT can boost the joint kilonova detection rates by detecting
more events at higher redshift. We also find some exceptions on the detection
numbers, e.g., $K$ samples with WFST in Table~\ref{ tab:kilonova detection rates
}. We suspect that this is attributed to the rounding and random errors in the
BNS population simulations, especially for a small number of events at lower
redshift. For a larger $t_{\mathrm{e}}$,  all the results should behave worse
when compared with the values listed in Table~\ref{ tab:kilonova detection rates
}.

\subsection{ Characteristics of Multi-messenger Events }
\label{ sec:Parameter Distributions }

As mentioned in \citet{Liu:2022mcd}, for BNSs in category (b), B-DEC could reach
a localization accuracy of $\Delta \Omega \lesssim 1\,\mathrm{deg}^2$ and timing
accuracy around $\Delta t_{c} \sim 0.1\,\mathrm{s}$, which is of great use to
multi-messenger astronomy. In this subsection, we first compare the $\Delta
\Omega$-$\Delta t_{c}$ and $z$-$\theta_{\mathrm{obs}}$ distributions of our
yearly early-warning samples in the LN model with $t_\mathrm{e} = 1\,\mathrm{d}$
for B-DEC and DO-OPT in Fig.~\ref{ fig:sGRB distributions }. Given that
different $\gamma$-ray detectors vary in the sGRB detection abilities (see
Sec.~\ref{ sec:sGRB Detection }), we mark all the GW early-warning events with
circles in different colors based on whether they can be observed by
$\gamma$-ray detectors.  Specifically, the majority of our early-warning samples
denoted by gray circles do not have joint sGRB detections. For the others, we
use green circles to denote joint detections that can only be achieved by SBSE,
while the blue ones are that can be detected by both SBSE and GECAM. All the
joint sGRB detections that can be achieved by the three $\gamma$-ray detectors
(FG, GACAM and SBSE) are denoted in red circles. In Fig.~\ref{ fig:sGRB
distributions }, we further plot  all the early-warning events with joint sGRB
detections in purple histograms (i.e., including the green, blue and red
circles). 

As clearly shown in Fig.~\ref{ fig:sGRB distributions }, one can achieve
$\lesssim 1\,\mathrm{deg}^2$ accuracy in localization for almost all events. It
is more than adequate for positioning every possible event considering the FoVs
of each $\gamma$-ray detector (cf. Sec.~\ref{ sec:Detection Rates }). The
$\Delta t_{\rm c}$ values of B-DEC and DO-OPT are both at $\sim
\mathcal{O}(0.1)\,{\rm s}$ level. Considering $t_\mathrm{e} = 1\,\mathrm{d}$ (or
an even larger $t_{\rm e}$), such a small timing accuracy will not affect the
overall early-warning detections in the wait-for scheme. We also find that the
joint sGRB detections are far more likely to be achieved with a lower redshift
and a smaller $\theta_{\mathrm{obs}}$ for the sources. The viewing angle
$\theta_{\mathrm{obs}}$ seems to be the dominating factor in detection
abilities, which can be explained well by the exponential decay in Eq.~(\ref{
eq:Gaussian-shaped jet profile }). 

\begin{figure*}
    \centering
    \includegraphics[width=8.3cm]{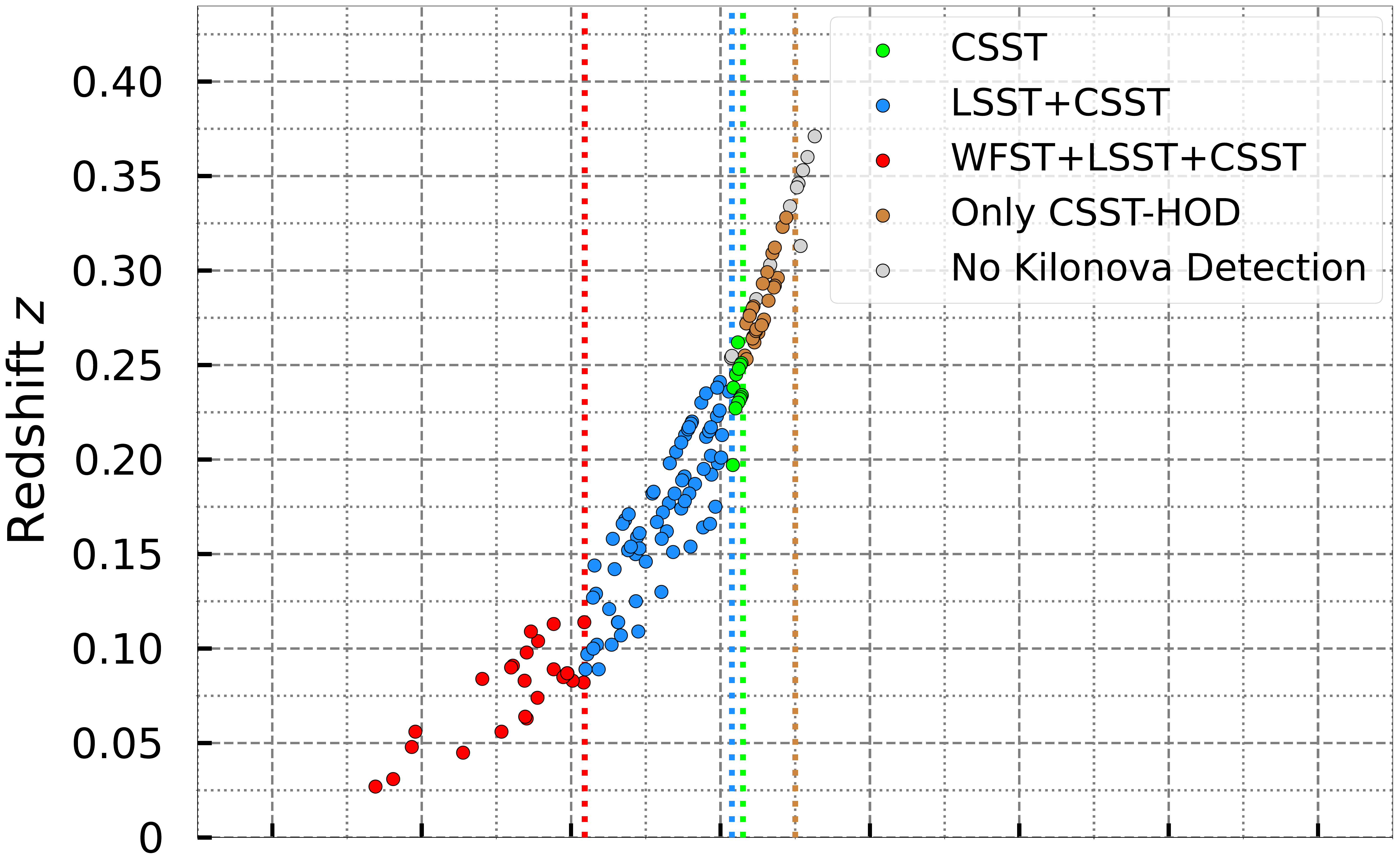}
    \hspace{1.0em}
    \includegraphics[width=8.15cm]{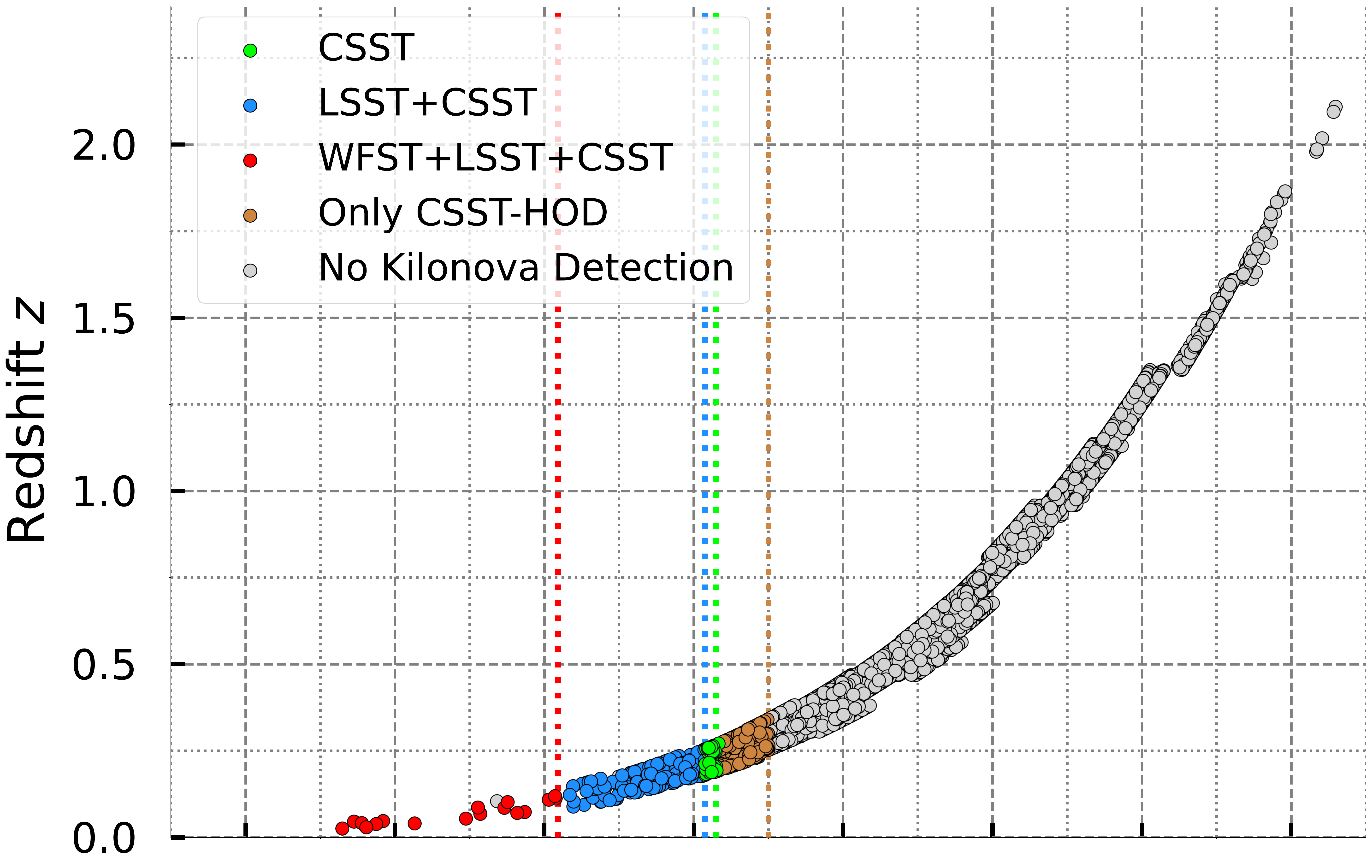}
    
    \vspace{-0.3em}
    
    \includegraphics[width=8.3cm]{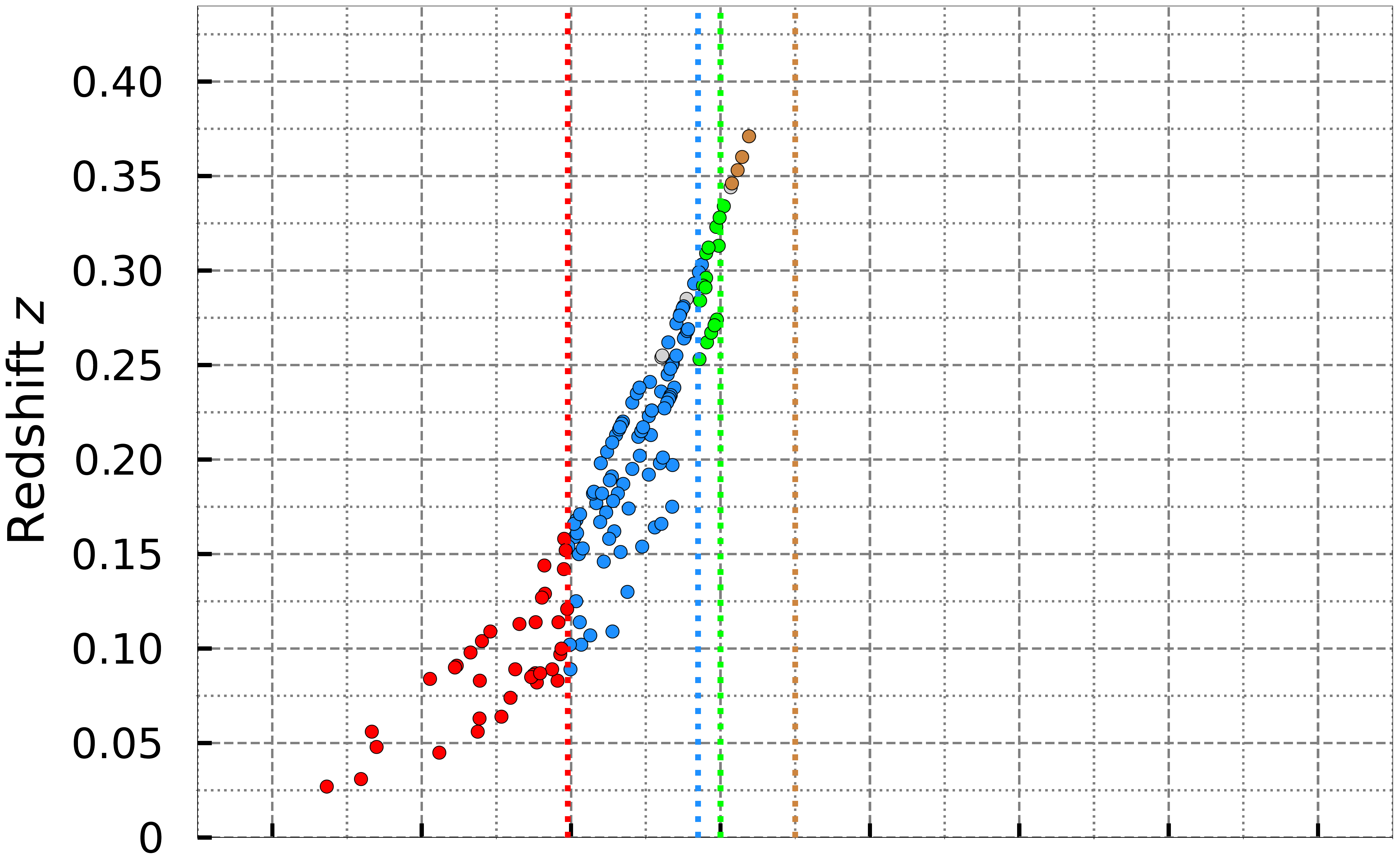}
    \hspace{1.0em}
    \includegraphics[width=8.15cm]{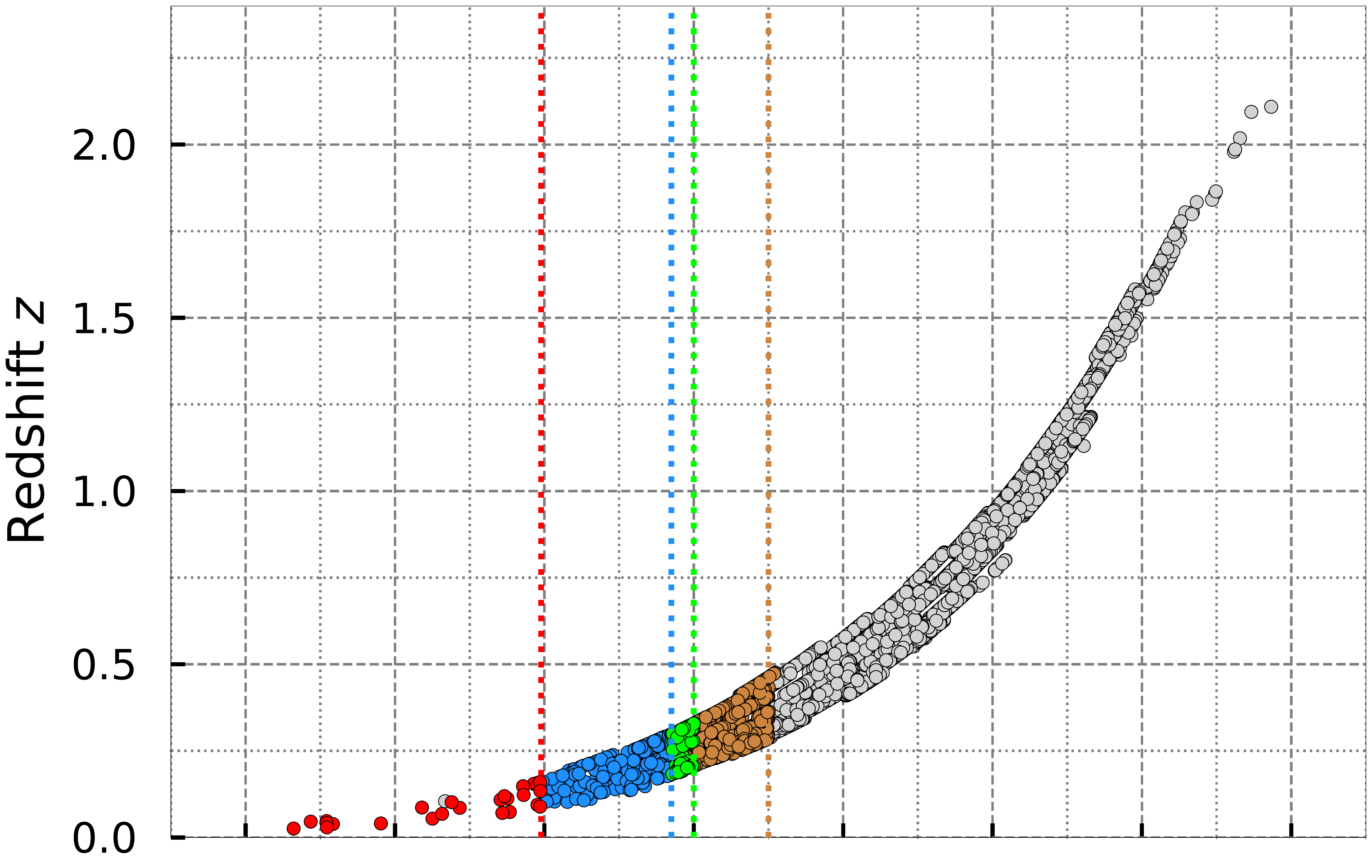}
    
    \vspace{-0.3em}
    
    \includegraphics[width=8.3cm]{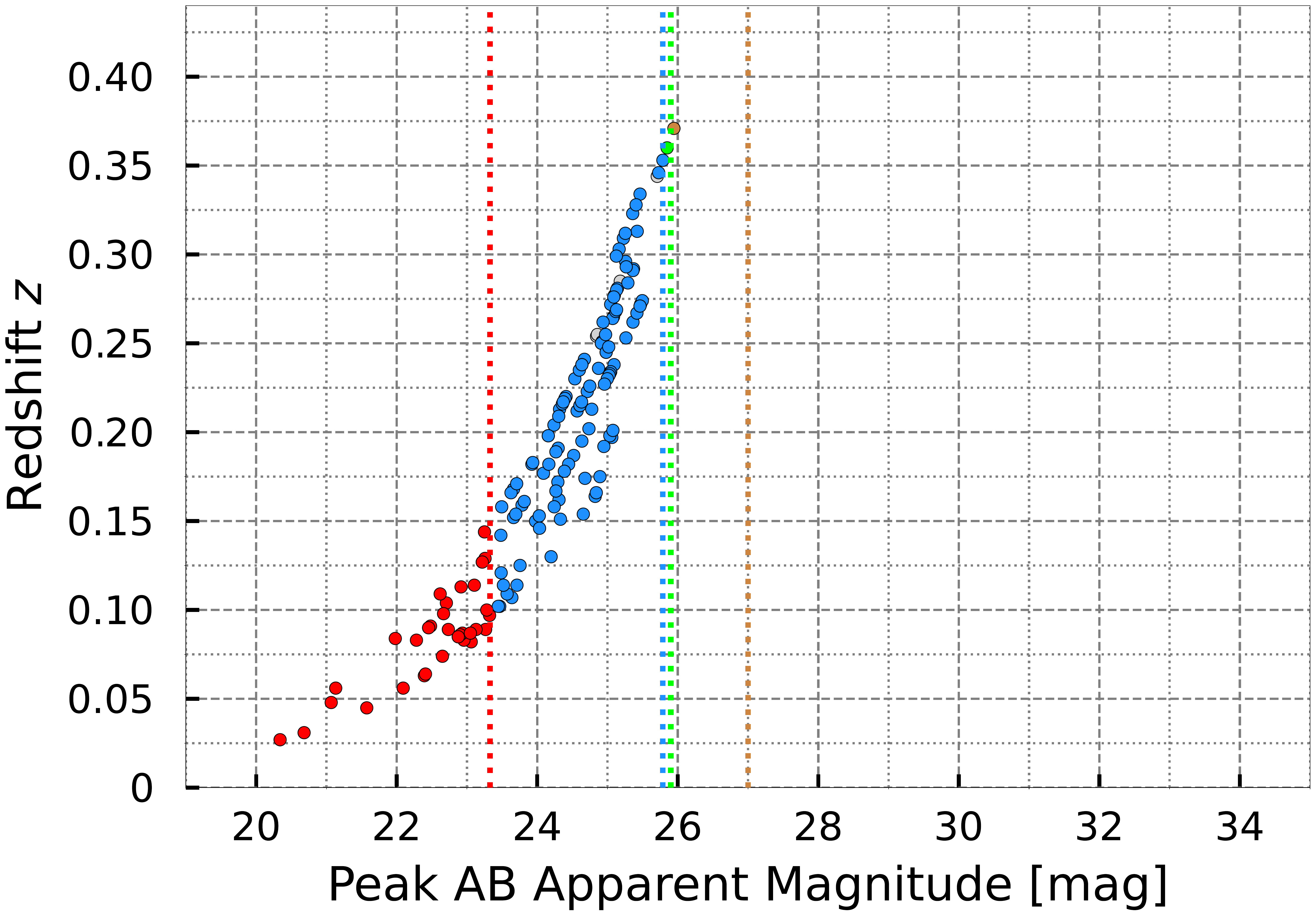}
    \hspace{1.0em}
    \includegraphics[width=8.15cm]{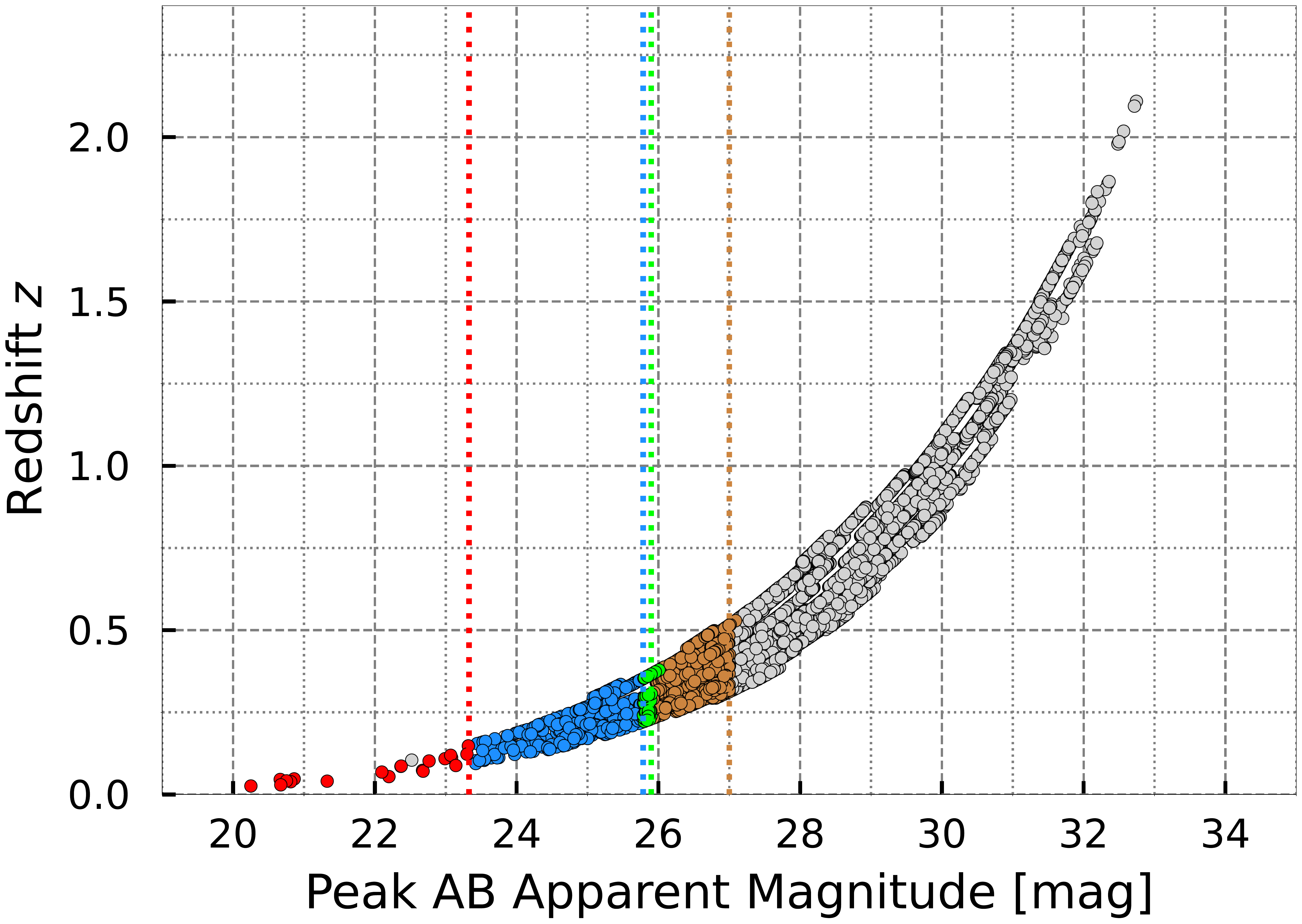}
    \caption{The relationship between the peak AB apparent magnitude of
    kilonovae and the redshift of the events for our yearly early-warning
    samples in the LN model with $t_\mathrm{e} = 1\,\mathrm{d}$ as an example.
    Left panels are for B-DEC while right panels are for DO-OPT. Top, middle,
    and bottom panels correspond to the detections in $\it{g}$, $\it{r}$, and
    $\it{i}$ bands, respectively. Dashed vertical lines with different colors
    denote the search limiting magnitude $m^*$ of corresponding survey missions
    (see Sec.~\ref{ sec:Kilonova Detection } and Table~\ref{ tab:Survey projects
    }). Circles with different colors represent different joint kilonova
    detections. } \label{ fig:kilonova distributions } 
\end{figure*}

For the multi-band kilonova detections, we illustrate in Fig.~\ref{ fig:kilonova
distributions } the relationship between the peak AB apparent magnitude of
kilonovae and the redshift of the events for our yearly early-warning samples in
the LN model with $t_\mathrm{e} = 1\,\mathrm{d}$. Given that there are few $GK$
events in our simulations (see Table~\ref{ tab:kilonova detection rates }), an
early-warning event with a joint GW-kilonova detection in Fig.~\ref{
fig:kilonova distributions } is defined as belonging to either the $\it{GK}$
samples or the $\it{K}$ samples. We again mark all the GW early-warning events
from B-DEC and DO-OPT with circles in different colors based on whether their
kilonovae can be detected by survey telescopes. Specifically, we use green
circles to denote joint kilonova detections that can only be achieved by CSST,
the most sensitive proposed survey telescope except the hypothetical CSST-HOD
(see Table~\ref{ tab:Survey projects }), while the blue ones are that to be
detected by both CSST and LSST. We use red circles to denote the events with
kilonovae  that can be detected by all three proposed survey missions (WFST,
LSST and CSST). The gray circles in Fig.~\ref{ fig:kilonova distributions }
represent the early-warning events without  joint kilonova detections. In
addition, we use the brown circles to denote events in isolation that can only
be seen by CSST-HOD due to its better $m^*$ values in each band (see Sec.~\ref{
sec:Kilonova Detection }).

Because B-DEC can only detect early-warning sources up to $z \simeq 0.45$, the
majority of its events are expected to have joint kilonova detections, at least
for the CSST. However, it seems rather difficult to detect most early-warning
events of DO-OPT due to their large redshifts, as shown in Fig.~\ref{
fig:kilonova distributions }. Considering the influence of redshift, we see that
the $\it{i}$-band observations perform better in Fig.~\ref{ fig:kilonova
distributions }, which can be an important consideration in the practical
observations, especially for events from DO-OPT.

As mentioned in Sec.~\ref{ sec:Detection Rates }, we no longer need to consider
the best searching strategy and FoV influence for our early-warning kilonova
detections like previous studies. It is because that with the accuracy of
$\Delta \Omega\lesssim 1\,\mathrm{deg}^2$ and $\Delta t_{\mathrm{c}}\lesssim
0.5\,\mathrm{s}$  shown in Fig.~\ref{ fig:sGRB distributions }, such a
sufficient early-warning time $t_\mathrm{e} = 1\,\mathrm{d}$ (or an even larger
$t_{\rm e}$) would allow us to locate the optical transients well with survey
projects in Table~\ref{ tab:Survey projects }. Supposing that there could be
HODs to cooperate with B-DEC and DO-OPT in the future---like the CSST-HOD and
SBSE-HOD-1 considered in this paper---we can detect more events and more distant
sources using multi-band observations as shown in Fig.~\ref{ fig:kilonova
distributions }. Especially, for the events with better location accuracy, more
delicate and informative observations are expected for other missions
with higher sensitivity rather than larger FoVs as those projects discussed in
our work, for an example, the James Webb Space Telescope with an FoV $\sim 100
\operatorname{arcmin}^{2}$ \citep{Gardner:2006ky, Yung:2021met}.

\section{ Conclusion }
\label{ sec:Conclusion }

In this work, we extend \citet{Liu:2022mcd} to propose a feasible wait-for
pattern as a novel detecting mode for various $\gamma$-ray detectors and optical
survey missions in the near future in synergy with space-borne decihertz GW
detectors. Such a methodology is based on the GW early-warning populations of
BNS mergers using two space-borne decihertz observatories.  The analysis is
performed using the method of Fisher matrix. With a sufficient early-warning
time $t_\mathrm{e} = 1\,\mathrm{d}$ (or an even larger $t_\mathrm{e}$), we point
out that one can prepare well in advance for the EM transients after the BNS
mergers and no longer needs to consider the FoV discounts and complex searching
strategies like in previous studies. We not only give quick assessments of
yearly joint sGRB detection rates for different $\gamma$-ray detectors and BNS
population models, but also report detailed analyses and results on kilonova
detections for several survey telescopes with different search limiting
magnitudes in $\it{g}$, $\it{r}$, and $\it{i}$ bands. We find that DO-OPT has
better performances than B-DEC as a whole during an assumed 4-yr mission time,
especially for the high-redshift events. Taking the LN population model with
$t_\mathrm{e} = 1\,\mathrm{d}$ as an example, we present the localization and
timing abilities of space-borne decihertz GW observatories and discuss the
redshift distributions for various EM and GW synergy observations. Furthermore,
given that there might be no overlapping observational time for proposed EM
projects and B-DEC/DO-OPT, we  have considered some HODs with better
sensitivities than the current designs in this work for multi-messenger
observations. Note that a specific ground-based EM project should consider the
discount due to its maximum detectable sky coverage and the Earth's rotation.
But for such a wait-for pattern, we suggest that this discount problem can be
partially solved if there are at least two survey telescopes on the different
sides of the Earth. In contrast to sGRBs, the relatively long duration of
kilonovae will also minish the discount.

While most research about multi-messenger observations of BNS mergers has
focused on the low-latency GW triggers and early warnings from ground-based GW
detectors \citep{Hooper:2011rb, Nitz:2018rgo, LIGOScientific:2019gag,
Sachdev:2020lfd, Nitz:2021pbr, Singh:2021zah, Magee:2022kkc, Borhanian:2022czq},
there are few studies on the realistic detections of BNS merger populations and
early warnings with  space-borne decihertz GW observatories \citep{Liu:2021dcr,
Liu:2022mcd}.  As we have demonstrated in this paper, space-borne decihertz
observatories could have a prominent advantage over ground-based GW detectors by
gathering enough information from the pre-merger stages of BNS mergers.
Decihertz GW observatories can also provide the location and time of the merger
in advance to the EM facilities. The $\mathcal{O}(0.01)\,\mathrm{deg}^2$
localization precision and $\mathcal{O}(0.1)\,\mathrm{s}$ timing accuracy
achieved by decihertz GW detectors will not only increase the joint detection
rates for the future EM follow-up projects, but also provide another
unprecedented opportunity for analyzing the early evolution of each detectable
system, such as the properties of pre- and post-merger precursors as suggested
by some observational results and theoretical works \citep{Troja:2010zm,
Li:2015pao, Gottlieb:2017pju, Bromberg:2017crh, Gottlieb:2019hfc,
Levinson:2019usn, Nakar:2019fza, Wang:2020vvr, Wang:2021msh, Sridhar:2020uez}.
We leave this to a future study.

In our Monte Carlo simulations, we have adopted a GRB\,170817A-like Gaussian jet
structure and AT\,2017gfo-like model to justify whether the simulated mergers
could be detected by various $\gamma$-ray detectors and optical survey missions. Our results can be used to provide meaningful references and helpful inputs for
upcoming EM follow-up projects by exploring the prospects for multi-messenger
early-warning detections.  We hope that more studies are carried out in the
future in order to give more precise predictions of sGRBs and kilonovae with
detailed analyses.

\section*{Acknowledgements}

It is a great pleasure to thank Jin-Ping Zhu for his useful advice and comments.
This work was supported by the National Natural Science Foundation of China
(11975027, 11991053, 11721303), the National SKA Program of China
(2020SKA0120300), the Max Planck Partner Group Program funded by the Max Planck
Society, and the High-Performance Computing Platform of Peking University. Y.K.
acknowledges the Hui-Chun Chin and Tsung-Dao Lee Chinese Undergraduate Research
Endowment (Chun-Tsung Endowment) at Peking University.

\section*{Data Availability}

The data underlying this paper will be shared on a reasonable request to the corresponding author.



\bibliographystyle{mnras}
\bibliography{refs} 



\bsp	
\label{lastpage}
\end{document}